\PassOptionsToPackage{usenames,dvipsnames,x11names}{xcolor}
\documentclass{JFM-FLM_Au}
\usepackage{url}
\usepackage{graphicx}
\usepackage{epstopdf, epsfig}
\usepackage{natbib}
\usepackage{paralist}
\usepackage[normalem]{ulem}
\usepackage[shortlabels]{enumitem}
\usepackage{CJK}
\usepackage{amssymb,euscript,latexsym,amsmath}
\usepackage{color}
\usepackage{soul}

\usepackage{epstopdf}
\usepackage{setspace}
\usepackage{chngpage}
\usepackage{autonum}
\usepackage{mdframed}   
\mdfsetup{%
linecolor=white,
backgroundcolor=gray!20,
}
\usepackage{ntheorem}
\theoremstyle{break}
\theoremheaderfont{\bfseries}
\newmdtheoremenv[%
linecolor=white,leftmargin=0,%
rightmargin=0,
backgroundcolor=gray!15,%
innertopmargin=0pt,%
ntheorem]{myprop}{Proposition}[section]
\usepackage{colortbl}
\usepackage[backgroundcolor=blue!10]{todonotes}

\newcommand{\mbx}{\boldsymbol{x}}
\newcommand{\mbf}{\boldsymbol{f}}
\newcommand{\mbu}{\boldsymbol{{u}}}

\newcommand{\mbv}{\boldsymbol{{v}}}

\definecolor{eggplant}{RGB}{126,32,181}

\title{Squirmers with arbitrary shape and slip: modeling, simulation, and optimization}

\author{Kausik Das\aff{1,\dag} , Hai Zhu\aff{2,\dag}, 
Marc Bonnet\aff{3}, 
\and Shravan Veerapaneni\aff{1}}
\affiliation{
\aff{1} Department of Mathematics, University of Michigan, Ann Arbor, MI, 48109 USA.
\aff{2}Flatiron Institute, Simons Foundation, New York, USA.
\aff{3}POEMS (CNRS, INRIA, ENSTA), ENSTA Paris, 91120 Palaiseau, France.}
\lefttitle{K. Das et al.}
\righttitle{Journal of Fluid Mechanics}
\corresau{Kausik Das, \email{kausik@umich.edu}}
\begin{document}
\begingroup
\let\thefootnote\relax
\footnotetext{\textsuperscript{\dag}These authors contributed equally to this work.}
\endgroup
\maketitle
\begin{abstract}
We consider arbitrary-shaped microswimmers of spherical topology and propose a framework for expressing their slip velocity in terms of tangential basis functions defined on the boundary of the swimmer using the Helmholtz decomposition. Given a time-independent slip velocity profile, we show that the trajectory followed by the microswimmer is a circular helix. We derive analytical expressions for the translational and rotational velocities of a prolate spheroid swimmer in terms of its Helmholtz decomposition modes and explore the effect of aspect ratio on these rigid body velocities. Then, for a given arbitrary swimmer shape of spherical topology, we investigate which slip profile minimizes the total power loss. A partial minimization is performed in which the direction of net motion of the swimmer is prescribed, followed by a global optimization procedure in which the best net motion direction is determined. The optimization results suggest that the competition between linear and rotational optimal motion is linked to symmetries in the shape of the microswimmer.

\end{abstract}
\section{Introduction}
\label{sc:intro}


The coordinated beating of cilia in organisms can create a directional fluid flow near cells. This key property of cilia has been adapted to perform a variety of functions. For example, in mammals, the epithelium is ciliated to move respiratory and cerebrospinal fluid \citep{Wan2020}, while  microswimmers such as \textit{Opalina} and \textit{Paramecium} instead use cilia for locomotion \citep{Lauga2009hydrodynamics}. 

When locomotion is of interest, the \textit{squirmer model} \citep{Blake1971spherical,Lighthill1952squirming} can be used to study the motion of these ciliates without having to model cilia individually. The model shows that small-amplitude, periodic deformations of a sphere lead to swimming in a low Reynolds number fluid \citep{Lighthill1952squirming}. \cite{Blake1971spherical} applied the model to ciliates by treating the tips of the beating cilia as a deformable envelope. The model can be specified to various degrees of detail. \cite{Ishikawa2020} made a distinction between the solid body of the swimmer and the ciliary envelope, applying a no-slip condition at the body surface and a shear stress at the tips of the cilia. On the other hand, \cite{pedley2016spherical} studied the more specific case of the \textit{steady squirmer} by assuming no radial displacements and a time-independent tangential slip velocity that represents the effect of the ciliary movement without creating an explicit ciliary layer.  Since then, the model has been extended to non-spherical squirmers such as ellipsoids \citep{kyoya2015shape} and dumbbells \citep{Ouyang2022}, as well as to non-Newtonian fluids \citep{pietrzyk2019flow,Gong2024,eastham2020axisymmetric}.

Researchers are interested in harnessing the locomotion strategies of biological microswimmers to create synthetic phoretic particles for various biomedical applications. For example, Janus particles have been synthesized that have a catalytic face on which a chemical reaction occurs, and an inert face on which no reaction occurs \citep{howse2007self}. This asymmetry generates a concentration gradient that leads to a slip velocity on the surface, making the squirmer model a natural framework for modeling these particles. Magnetically-mediated Janus particles have been used to transport thrombin and microporous starch to wounds to combat severe bleeding \citep{LiJanus}. Recent work has focused on studying non-spherical Janus particle geometries such as dimers and spheroids \citep{Michelin2017}, but there is now a growing interest in deformable swimmers as they possess extra degrees of freedom to swim efficiently through narrow and viscous channels \citep{pramanik2024nature}. 

Some swimmers have been developed to replicate ciliate swimming techniques, while others, known as helical swimmers, swim by generating chirality using a body twist \citep{pramanik2024nature}. While simple squirmer models use an axisymmetric slip velocity, this does not capture the chirality present in helical swimmers and some ciliates. In fact, chirality is naturally present in many microorganisms and can also result from imperfections in the fabrication of synthetic swimmers \citep{samatas2023hydrodynamic}. The \textit{chiral squirmer model} \citep{Buradachiral} imposes a non-axisymmetric slip to replicate the chirally asymmetric paths followed by swimmers such as Marine Zooplankton. So far, researchers have studied the interactions between spherical chiral squirmers \citep{samatas2023hydrodynamic, maity2022near}, spherical chiral squirmers in a non-Newtonian fluid \citep{kobayashi2025viscotaxis}, and spring-connected chiral squirmers \citep{hu2025cargo}. Non-spherical chiral squirmers remain relatively unexplored, motivating the need for theoretical and computational studies of swimmers with \textit{arbitrary} shape and slip to inform the design and synthesis of new artificial swimmers.

Unlike the axisymmetric case, in which swimmers move in a straight line, it has been observed both experimentally and numerically that chiral swimmers rotate and move in helical paths \citep{Friedrichhelix,Crenshawhelix,Ishikawa2019, samatas2023hydrodynamic,kobayashi2025viscotaxis}. The helices allow organisms to orient to stimuli using periodic signalling, thereby avoiding reliance on a biased random walk \citep{Crenshawhelix,ambrosi2018}. 
There have been previous studies in the biology community on the helical motion of generic organisms -- \cite{crenshaw1993orientationII} determined that the axis of the helix is determined by the angular velocity, so that organisms change their angular velocity, rather than their translational velocity, to change direction. The study of helical trajectories in the context of the squirmer model has only recently gained traction \citep{ambrosi2018, kobayashi2025viscotaxis}. For example, \cite{ambrosi2018} proved that a microswimmer with periodically beating cilia follows a discrete circular helix trajectory by showing that the discrete curvature and torsion were constants. 
In our work, we provide a simpler proof of this result for squirmers with time-independent slip, avoiding the differential-geometric machinery of
\cite{crenshaw1993orientationII} and \cite{ambrosi2018}.

In the design of bioinspired microswimmers, optimizing for efficiency is a natural objective. Previous work by \cite{Giulianioptim} optimized a robotic swimmer consisting of a circular head and a rotating helical flagellum. They found the optimal flagellum length to maximize energy efficiency and displacement. \cite{Gidituriflagella} showed that these flagellated swimmers could be mapped to slip velocity-driven swimmers by discretizing the slip-driven swimmer into a collection of spheres that produces similar swimming velocities and flow fields. Thus, optimization results for slip-driven squirmers are applicable to other types of swimmers as well. \cite{Michelin2010efficiency} found that time-periodic metachronal beating of cilia optimized the swimming efficiency of spherical squirmers, while \cite{vilfan2012optimal} and \cite{guo2021optimal} found the optimal shape and slip of axisymmetric microswimmers to minimize power loss. There has also been recent progress in identifying the swimmer types (pusher/puller/neutral) of optimal axisymmetric squirmers \citep{zayed2025swimmer}. To optimize chiral squirmers, these optimization results need to be extended to non-axisymmetric shape and slip. Since arbitrary-shape squirmers can rotate, additional degrees of freedom are introduced, and thus the objective functions and constraints used for the optimization of axisymmetric squirmers must be modified. \cite{nasouri2021minimum} defined a new measure of efficiency for microswimmers of arbitrary shape. 

In this work, we optimize arbitrary-shape swimmers by finding the slip profile that minimizes power loss given a prescribed direction of net motion. We then perform a global optimization over the direction of net motion.
A prerequisite for such optimization is a systematic framework for defining slip velocities on arbitrary shapes.  For spherical squirmers, many works use only two modes of the slip velocity -- one corresponding to the swimming velocity and the other corresponding to the suspension viscosity \citep{Ishikawareview}, while others use an infinite series \citep{Ghoseseries,Buradachiral}. In this work, we extend these results to arbitrary shapes of spherical topology by making use of the Helmholtz decomposition to expand the tangential slip velocity into spherical harmonic modes. 

To maintain focus on the physical insights, we present the main framework and results here, deferring detailed mathematical derivations to our companion paper \citep{B-2025-06}. In particular, in this paper we focus on the forward problem with prescribed slip as well as the results of the slip optimization, leaving the derivation and discussion of the optimization algorithm, as well as implementation details for the companion paper. 

This paper is organized as follows. In Section \ref{sc:model}, we discuss the governing equations and boundary conditions for the problem as well as a decomposition for the slip velocity on arbitrary surfaces in terms of tangential basis functions. In Section \ref{sc:helix}, we provide a proof of a result regarding the helical trajectories of microswimmers with time-independent slip in free space and describe a method to solve for the rigid body velocities that determine the swimmer trajectories. Next, in Section \ref{sc: spheroid}, for the special case of a prolate spheroid, we derive analytical expressions for the resulting translational and rotational velocities, and explore the effect of aspect ratio on these rigid body velocities. Finally, in Section \ref{sc:optim} we outline the objective functions and constraints for both the partial slip optimization in which the direction of net motion is prescribed, and the global optimization in which the best net motion direction is determined, followed by optimal solutions for various shapes in Section \ref{sc:results} and conclusions in Section \ref{sc:conclusion}.






\section{Mathematical model}
\label{sc:model}

Consider a three-dimensional microswimmer of arbitrary shape with a boundary $\Gamma$ suspended in an unbounded viscous fluid domain. In the low Reynolds number limit, the fluid surrounding the microswimmer is governed by the incompressible Stokes equations:
\begin{align}
\centering
\label{eq:Stokes}
-\boldsymbol{\nabla} p + \nabla^2 \mbu = \boldsymbol{0}  \quad\text{and}\quad
\boldsymbol{\nabla} \boldsymbol{\cdot} \mbu = 0,
\end{align} 
where $\mbu$ is the fluid velocity and $p$ is the pressure. Note that we have nondimensionalized the equations for convenience. In addition to the Stokes equations, we have boundary conditions in the far field and on the swimmer surface, given by
\begin{align}
\centering
\label{eq:Stokesbc}
\lim_{|\boldsymbol{x}|\rightarrow\infty}\mbu = \boldsymbol{0} \quad\text{and}\quad
  \mbu = \mbv^{\textrm{S}} + \mbv^{\textrm{R}} \quad\text{on}\quad \Gamma,   
\end{align}
where $\mbv^{\textrm{S}}$ is a tangential slip velocity and the rigid body velocity field $\boldsymbol{v}^\textrm{R} := \boldsymbol{U} + \boldsymbol{\Omega} \times \mbx $, with $\boldsymbol{U}$ and $\boldsymbol{\Omega}$ denoting the translational and angular velocities of the swimmer, respectively. In the absence of external stimuli, the total force and torque experienced by the swimmer must vanish, so an additional condition to be imposed is the following: 
\begin{align}
\label{eq: forcetorquefree}
\int_{\Gamma} \mbf \, \mathrm{d} S &= \boldsymbol{0}\quad\text{and}\quad \int_{\Gamma} \mbx \times \mbf \, \mathrm{d} S = \boldsymbol{0},
\end{align}
where $\mbf$ is the surface traction (force density).

A more general formulation is needed to extend existing definitions of the slip velocity $\mbv^{\textrm{S}}$ for spherical and axisymmetric squirmers to general three-dimensional shapes. We can use the Helmholtz decomposition to write any vector field tangent to $\Gamma$ as the sum of a surface curl-free part involving the surface gradient $\boldsymbol{\nabla}_\Gamma$ and a surface divergence-free part involving the surface curl  $\boldsymbol{\nabla}\times_\Gamma$, so that 
\begin{align}
\label{eq: Helmholtz}
 \boldsymbol{v}^\textrm{S} = \boldsymbol{\nabla}_\Gamma \Phi + \boldsymbol{\nabla}\times_\Gamma \Psi = \boldsymbol{\nabla}_\Gamma \Phi + \boldsymbol{\nabla}_\Gamma \Psi \times \boldsymbol{n}
\end{align}
for some scalar-valued functions $\Phi$ and $\Psi$, where $\boldsymbol{n}$ is the unit outward normal vector~\citep[eq.~5.6.24]{nedelec:book}.
Assuming spherical topology, we can express $\Phi$ and $\Psi$ in the spherical harmonics basis (defined on $\Gamma$ via a smooth mapping from the unit sphere) to expand $\mbv^{\textrm{S}}$ in the form
\begin{align}
\label{eq: slipdecomp}
            \mbv^{\textrm{S}}(\theta,\phi) = \sum_{n=1}^{\infty}\sum_{m=-n}^{n} a_n^m\boldsymbol{\nabla}_\Gamma Y_n^m (\theta,\phi) + \sum_{n=1}^{\infty}\sum_{m=-n}^{n} b_n^m\boldsymbol{\nabla}_\Gamma Y_n^m (\theta,\phi) \times \boldsymbol{n} (\theta,\phi),
\end{align}
where $\theta\!\in\![0,\pi]$ and $\phi\!\in\![0,2\pi]$ are the usual polar and azimuthal angular spherical coordinates respectively, $Y_n^m$ is the standard scalar spherical harmonic function of degree $n$ and order $m$, $a_n^m$ and $b_n^m$ are complex coefficients representing mode amplitudes. This generalizes the decomposition used by \cite{Buradachiral} for spheres, which can be obtained by setting $\boldsymbol{n} = \boldsymbol{e}_r$ in \eqref{eq: slipdecomp}. For spherical squirmers with axisymmetric slip, only the $m = 0$ terms are retained so that
\begin{align}
\label{eq: slipdecompaxisym}
    \mbv^{\textrm{S}}(\theta,\phi) = -\sum_{n=1}^{\infty} \frac{a_n^0}{a} \sqrt{\frac{2n+1}{4 \pi}} P_n'(\cos \theta) \sin \theta \boldsymbol{e}_\theta  + \sum_{n=1}^{\infty} \frac{b_n^0}{a} \sqrt{\frac{2n+1}{4 \pi}} P_n'(\cos \theta) \sin \theta \boldsymbol{e}_\phi
\end{align}
where $P_n$ is the Legendre polynomial of degree $n$ and $a$ is the sphere radius. This agrees with the form of the expansion seen in \cite{pak2014generalized}. Typically, studies include only the first two squirming modes of the above expansion, such that 
\begin{align}
\label{eq: slipdecomptwoparam}
    \mbv^{\textrm{S}}(\theta,\phi)  = -\sqrt{\frac{3}{4 \pi}}\frac{a_1^0}{a} \Bigl[\, \sin \theta + \sqrt{15}(a_2^0/a_1^0) \sin \theta \cos \theta \,\Bigr] \boldsymbol{e}_\theta,
\end{align}
as seen in \cite{Ishikawareview}. The two modes define the translational velocity $U_z = -a_1^0 \sqrt{\frac{3}{\pi}} \frac{1}{3a}$ and the squirmer parameter $\beta = \sqrt{15}a_2^0/a_1^0$, which classifies swimmers as pushers and pullers. However, as we will see in the following sections, including non-axisymmetric modes allows for more generalized motion in the $x$- and $y$-directions. While including only axisymmetric modes leads to straight line motion along the axis of symmetry, the motion due to a general slip velocity with non-axisymmetric modes is in fact helical \citep{kobayashi2025viscotaxis}, as we will show in the next section. 


    
\section{Helical orbits}
\label{sc:helix}

\vspace{0.1in}
In this section, we show that microswimmers in free space with a time-independent slip velocity follow circular helical trajectories. To begin, observe that when the prescribed slip $\boldsymbol{v}^S$ is time-independent,
the Stokes equations and boundary conditions are time-independent, and by uniqueness of the solution to Stokes equations, 
the rigid body velocities $\boldsymbol{U}$ and $\boldsymbol{\Omega}$ are both constants when viewed in a coordinate frame attached to the swimmer. 
%
%
%
Since they are constants, the swimmer trajectory can be obtained by computing $\boldsymbol{U}$ and $\boldsymbol{\Omega}$ at $t = 0$ using one forward solve and then evolving the centroid position using the ordinary differential equation (ODE) $\dot{\boldsymbol{x}}^c = R(t) \boldsymbol{U}$, where $R(t)$ is a rotation matrix that rotates the body frame velocity $\boldsymbol{U}$ to obtain the centroid velocity in the lab frame. If $\boldsymbol{\Omega} = \boldsymbol{0}$, then $R(t) = I$ and the ODE reduces to $\dot{\boldsymbol{x}}^c = \boldsymbol{U}$, giving a straight line trajectory, which is a degenerate helix of radius zero.  

For $\boldsymbol{\Omega} \neq  \boldsymbol{0}$, using Rodrigues' rotation formula, $R(t)$ can be evaluated as $R(t) = I + \sin (|\boldsymbol{\Omega}|t) M + [1- \cos (|\boldsymbol{\Omega}|t)]M^2$, where 
\begin{align}
M = \frac{1}{|\boldsymbol{\Omega}|}
\begin{bmatrix}
    0 & -\Omega_z & \Omega_y \\
    \Omega_z & 0 & - \Omega_x \\
    -\Omega_y & \Omega_x & 0
\end{bmatrix}
\end{align}
\citep{CORONA2017504}. Integrating the ODE and using the property that $M \boldsymbol{v}  = \frac{\boldsymbol{\Omega}} {|\boldsymbol{\Omega}|} \times \boldsymbol{v}$ for any vector $\boldsymbol{v}$, we obtain a parametrization for the trajectory of the centroid, given by 

\begin{align}
\boldsymbol{x}^c(t) = \boldsymbol{x}_0 + \frac{\boldsymbol{\Omega} \boldsymbol{\cdot} \boldsymbol{U}}{|\boldsymbol{\Omega}|^2} \boldsymbol{\Omega}  t  - \frac{\boldsymbol{\Omega} \times \boldsymbol{U}}{|\boldsymbol{\Omega}|^2} \cos (|\boldsymbol{\Omega}|t)  -\frac{\boldsymbol{\Omega} \times (\boldsymbol{\Omega} \times \boldsymbol{U})}{|\boldsymbol{\Omega}|^3}\sin (|\boldsymbol{\Omega}|t).
\label{eq: altparam} \end{align}
where $\boldsymbol{x}_0$ is a constant vector. 

In its most general form where $\boldsymbol{\Omega} \boldsymbol{\cdot} \boldsymbol{U} \neq 0$ and $\boldsymbol{\Omega} \times \boldsymbol{U} \neq \boldsymbol{0}$, \eqref{eq: altparam} can be written as 

\begin{align}
\boldsymbol{x}^c(t) = \boldsymbol{x}_0 + \frac{\boldsymbol{\Omega} \boldsymbol{\cdot} \boldsymbol{U}}{|\boldsymbol{\Omega}|} t \boldsymbol{h}_1 -  \frac{|\boldsymbol{\Omega}\times \boldsymbol{U}|}{|\boldsymbol{\Omega | }^2}\cos (|\boldsymbol{\Omega}|t) \boldsymbol{h}_2 -\frac{|\boldsymbol{\Omega}\times \boldsymbol{U}|}{|\boldsymbol{\Omega | }^2}\sin (|\boldsymbol{\Omega}|t) \boldsymbol{h}_3 
\label{eq: helix}\end{align}
 where
 \begin{align}
\boldsymbol{h}_1 = \frac{\boldsymbol{\Omega}}{|\boldsymbol{\Omega}|}, \quad \boldsymbol{h}_2 = \frac{\boldsymbol{\Omega} \times \boldsymbol{U}}{|\boldsymbol{\Omega} \times \boldsymbol{U}|},\quad \boldsymbol{h}_3 = \frac{\boldsymbol{\Omega} \times (\boldsymbol{\Omega} \times \boldsymbol{U})}{|\boldsymbol{\Omega} \times (\boldsymbol{\Omega} \times \boldsymbol{U})|}.
\end{align}
By the properties of the cross product, $\{\boldsymbol{h}_1, \boldsymbol{h}_2,\boldsymbol{h}_3\}$ form an orthonormal set. Thus \eqref{eq: helix} parametrizes a circular helix with radius $|\boldsymbol{\Omega}\times \boldsymbol{U}|/|\boldsymbol{\Omega | }^2$ and pitch $2 \pi |\boldsymbol{\Omega} \boldsymbol{\cdot} \boldsymbol{U}|/|\boldsymbol{\Omega}|^2$, in agreement with the result from \cite{Buradachiral} for spherical chiral squirmers. The helix consists of linear motion along the helix axis $\boldsymbol{h}_1$ and rotational motion in the plane spanned by $\boldsymbol{h}_2$ and $\boldsymbol{h}_3$. The direction of the helix axis is defined by the direction of $\boldsymbol{\Omega}$; this agrees with the result found in \cite{crenshaw1989kinematics}.

We also note that straight line (helix with zero radius) and circular (helix with zero pitch) trajectories can be obtained as special cases of \eqref{eq: altparam} when  $\boldsymbol{\Omega} \times \boldsymbol{U} = \boldsymbol{0}$ and $\boldsymbol{\Omega} \boldsymbol{\cdot} \boldsymbol{U} = 0$, respectively. The various classes of trajectories are summarized in Table \ref{table: dotcrossclassfication} and match the observations made by \cite{Crenshawhelix} and \cite{Buradachiral}.

\begin{table}
\centering
\renewcommand{\arraystretch}{1.25}
\begin{tabular}{>{\raggedright\arraybackslash}m{4cm} | >{\centering\arraybackslash}m{4cm} >{\centering\arraybackslash}m{4cm} }
  & $\boldsymbol{\Omega} \boldsymbol{\cdot} \boldsymbol{U} = 0$ & $\boldsymbol{\Omega} \boldsymbol{\cdot} \boldsymbol{U} \neq 0$ \\
  \hline
  $\boldsymbol{\Omega} \times \boldsymbol{U}= \boldsymbol{0}$ & Point & Straight line \\
  \hline
  $\boldsymbol{\Omega} \times \boldsymbol{U} \neq \boldsymbol{0}$ & Circle & Circular helix \\
\end{tabular}
\caption{Classification of helical trajectories based on the values of $\boldsymbol{\Omega} \boldsymbol{\cdot} \boldsymbol{U}$ and $\boldsymbol{\Omega} \times \boldsymbol{U}$ with $\boldsymbol{\Omega} \neq \boldsymbol{0}$.}
\label{table: dotcrossclassfication}
\end{table}


These helical trajectories for microorganisms have been studied extensively in the zoology and biology communities -- see the numerous works by \cite{crenshaw1989kinematics,Crenshawhelix,crenshaw1993orientation,crenshaw1993orientationIII,crenshaw1993orientationII}. In the fluid dynamics community, works such as \cite{Buradachiral} include the helical trajectory result without proof. While initial works assumed fewer than six degrees of freedom by restricting $\boldsymbol{U}$ to one component \citep{crenshaw1989kinematics}, later works allowed for a full six degrees of freedom \citep{crenshaw1993orientation}, just as we see in our result. Our proof also avoids dealing with the Frenet-Serret frames directly, unlike the treatments seen in the previously mentioned papers. While \cite{ambrosi2018} proved that microswimmers with periodic slip velocities follow discrete circular helix trajectories, our proof can be viewed as a continuous analogue of theirs in the case of time-independent slip velocities. In addition, while \cite{ambrosi2018} produced expressions for helix radius and pitch involving projections of displacement and angles of rotation, our result more conveniently describes these  helix properties in terms of the rigid body velocities $\boldsymbol{U}$ and $\boldsymbol{\Omega}$. 

The rigid body velocities are quantities of importance as they completely determine the swimmer trajectory. To determine these velocities, previous works have used the Lorentz reciprocal theorem to produce integral formulas for $\boldsymbol{U}$ and $\boldsymbol{\Omega}$ for geometries such as spheres and ellipsoids without having to solve the Stokes equations \citep{Masoud_Stone_2019reciprocal, stone1996propulsion}. These works utilized known analytical results for Stokes flow past these shapes. For arbitrary shape swimmers that can translate and rotate, \cite{Masoud_Stone_2019reciprocal} showed that 
\begin{align}
  \langle \boldsymbol{U}, \widehat{\boldsymbol{f}} \rangle_\Gamma + \langle \boldsymbol{\Omega} , \boldsymbol{x} \times \widehat{\boldsymbol{f}} \rangle_\Gamma = -\langle \boldsymbol{v}^\textrm{S}, \widehat{\boldsymbol{f}} \rangle_\Gamma 
  \label{eq: reciprocal}
\end{align}
where $\widehat{\boldsymbol{f}}$ is the traction when the Stokes equations \eqref{eq:Stokes} are subject to the no-slip boundary condition $\boldsymbol{u} = \widehat{\boldsymbol{U}} + \widehat{\boldsymbol{\Omega}} \times \boldsymbol{x}$ on $\Gamma$, and $\langle \boldsymbol{\cdot}, \boldsymbol{\cdot} \rangle_\Gamma$ is the $L^2$ inner product of vector fields on $\Gamma$. The six components of $\boldsymbol{U}$ and $\boldsymbol{\Omega}$ must be determined, but \eqref{eq: reciprocal} provides only one scalar equation. To obtain six equations instead, we can set $\widehat{\boldsymbol{U}} = \boldsymbol{e}_\ell$, $\widehat{\boldsymbol{\Omega}} = \boldsymbol{0}$ for $\ell = 1,2,3$ and $\widehat{\boldsymbol{U}} = \boldsymbol{0},$ $\widehat{\boldsymbol{\Omega}} = \boldsymbol{e}_{\ell-3}$ for $\ell = 4,5,6$. Letting $\boldsymbol{f}^\textrm{R}_\ell$, $\ell = 1,\dots,6$ denote the tractions associated with these six auxiliary problems, we obtain a six-dimensional linear system 
\begin{align}
\begin{bmatrix}
\left(\int_\Gamma \boldsymbol{f}_1^\textrm{R} \, \mathrm{d}S \right)^T & \left(\int_\Gamma \boldsymbol{x} \times \boldsymbol{f}_1^\textrm{R} \, \mathrm{d}S \right)^T \\
\vdots & \vdots \\
\left(\int_\Gamma \boldsymbol{f}_6^\textrm{R} \, \mathrm{d}S \right)^T & \left(\int_\Gamma \boldsymbol{x} \times \boldsymbol{f}_6^\textrm{R} \, \mathrm{d}S \right)^T
\end{bmatrix}
\begin{bmatrix}
\boldsymbol{U} \\
\boldsymbol{\Omega}
\end{bmatrix}
 = 
 \begin{bmatrix}
- \int_\Gamma \boldsymbol{v}^\textrm{S} \boldsymbol{\cdot} \boldsymbol{f}_1^\textrm{R} \, \mathrm{d}S \\
\vdots \\
- \int_{\Gamma}\boldsymbol{v}^\textrm{S} \boldsymbol{\cdot} \boldsymbol{f}_6^\textrm{R} \, \mathrm{d}S
 \end{bmatrix},
 \label{eq: matrixeqn}
\end{align}
the matrix in the left-hand side being the matrix $\mathbf{C}$ of~\cite{B-2025-06}.
For each auxiliary problem, the tractions $\boldsymbol{f}^\textrm{R}_\ell$ can be computed either analytically for simple shapes (e.g., sphere) or using a numerical solver for Stokes equations, completely determining the matrix on the left hand side of \eqref{eq: matrixeqn}. That system can then be solved to determine the components of $\boldsymbol{U}$ and $\boldsymbol{\Omega}$ for any prescribed slip velocity $\boldsymbol{v}^\textrm{S}$.

\section{Analysis of first-order squirming modes on prolate spheroidal microswimmers}
\label{sc: spheroid}
%
The reciprocal theorem provides a powerful framework for determining rigid body velocities from a prescribed slip velocity. In this section, we apply it to derive analytical formulas for the rigid body velocities of a prolate spheroidal squirmer with equatorial radius $a$ and polar radius $b > a$ when only the $n = 1$ modes are retained in the slip decomposition \eqref{eq: slipdecomp}:
\begin{equation}
\label{eq: slipdecompfirstorder}
            \mbv^{\textrm{S}}(\theta,\phi) = \sum_{m=-1}^{1} a_1^m\boldsymbol{\nabla}_\Gamma Y_1^m + \sum_{m=-1}^{1}b_1^m\boldsymbol{\nabla}_\Gamma Y_1^m\times \boldsymbol{n}.
\end{equation}
To ensure that $\mbv^{\textrm{S}}$ is real, the coefficients are related by $a_1^{-1} = -\overline{a_1^1}$, $a_1^0 = \overline{a_1^0}$, $b_1^{-1} = -\overline{b_1^1}$, and $b_1^0 = \overline{b_1^0}.$ From \cite{Masoud_Stone_2019reciprocal}, the reciprocal theorem can be used to determine the translational and angular velocities as 
\begin{align} \label{eq:Uspheroid}
\boldsymbol{U} &= - \frac{1}{4 \pi a^2 b} \int_{\Gamma}(\boldsymbol{n} \boldsymbol{\cdot} \boldsymbol{x}) \mbv^{\textrm{S}} \, \mathrm{d}S, \\
\label{eq:Omegaspheroid}
\boldsymbol{\Omega} &= - \frac{3}{4 \pi a^2 b} \int_{\Gamma} (\boldsymbol{n} \boldsymbol{\cdot} \mbx) \mathcal{D}  (\mbx \times \mbv^{\textrm{S}}) \, \mathrm{d}S,
\end{align}
where $\mathcal{D} = \mathrm{diag}\big(1/(a^2 + b^2), 1/(a^2 + b^2), 1/(2a^2)\big)$.
%
By evaluating the integrals in \eqref{eq:Uspheroid} and \eqref{eq:Omegaspheroid}, the rigid body velocities can be expressed in terms of shape-dependent functions $f_\parallel, f_\perp$ (translation) and $g_\parallel, g_\perp$ (rotation):
\begin{align} \label{eq: spheroidrbv}
U_x &= \mathrm{Re}(a_1^1) f_\perp, & U_y &= -\mathrm{Im}(a_1^1) f_\perp, & U_z &= -a_1^0 f_\parallel, \\
\Omega_x &= \mathrm{Re}(b_1^1) g_\perp, & \Omega_y &= -\mathrm{Im}(b_1^1) g_\perp, & \Omega_z &= b_1^0 g_\parallel,
\end{align}
where $\xi = b/a > 1$ is the aspect ratio, and the geometric factors are
\begin{align}
f_\perp &= \frac{1}{2a}\sqrt{\frac{3}{2 \pi}} \left[ \frac{\xi^2 -2}{\xi^2 -1} + \frac{ \xi \cosh^{-1}\xi}{( \xi^2 -1)^{3/2}} \right], \qquad
f_\parallel = \frac{1}{2a}\sqrt{\frac{3}{\pi}} \left[ \frac{\xi}{\xi^2 - 1} - \frac{ \cosh^{-1}\xi}{(\xi^2-1)^{3/2}} \right], \\
g_\perp &= \frac{3\xi}{a^2(1+\xi^2)}\sqrt{\frac{3}{2 \pi}} \left[ \frac{(3 \xi^2-2) \sec^{-1}\xi}{4(\xi^2 - 1)^{3/2}} -\frac{1}{4(\xi^2-1)}\right], \\
g_\parallel &= \frac{3}{a^2}\sqrt{\frac{3}{\pi}} \left[ \frac{(2 - \xi^2)\sec^{-1}\xi}{8(\xi^2 -1)^{3/2}} - \frac{1}{8(\xi^2 - 1)}  \right].
\end{align}

\begin{figure}[t]
    \centerline{\includegraphics[trim=0 0 0 0, clip, width=\textwidth]{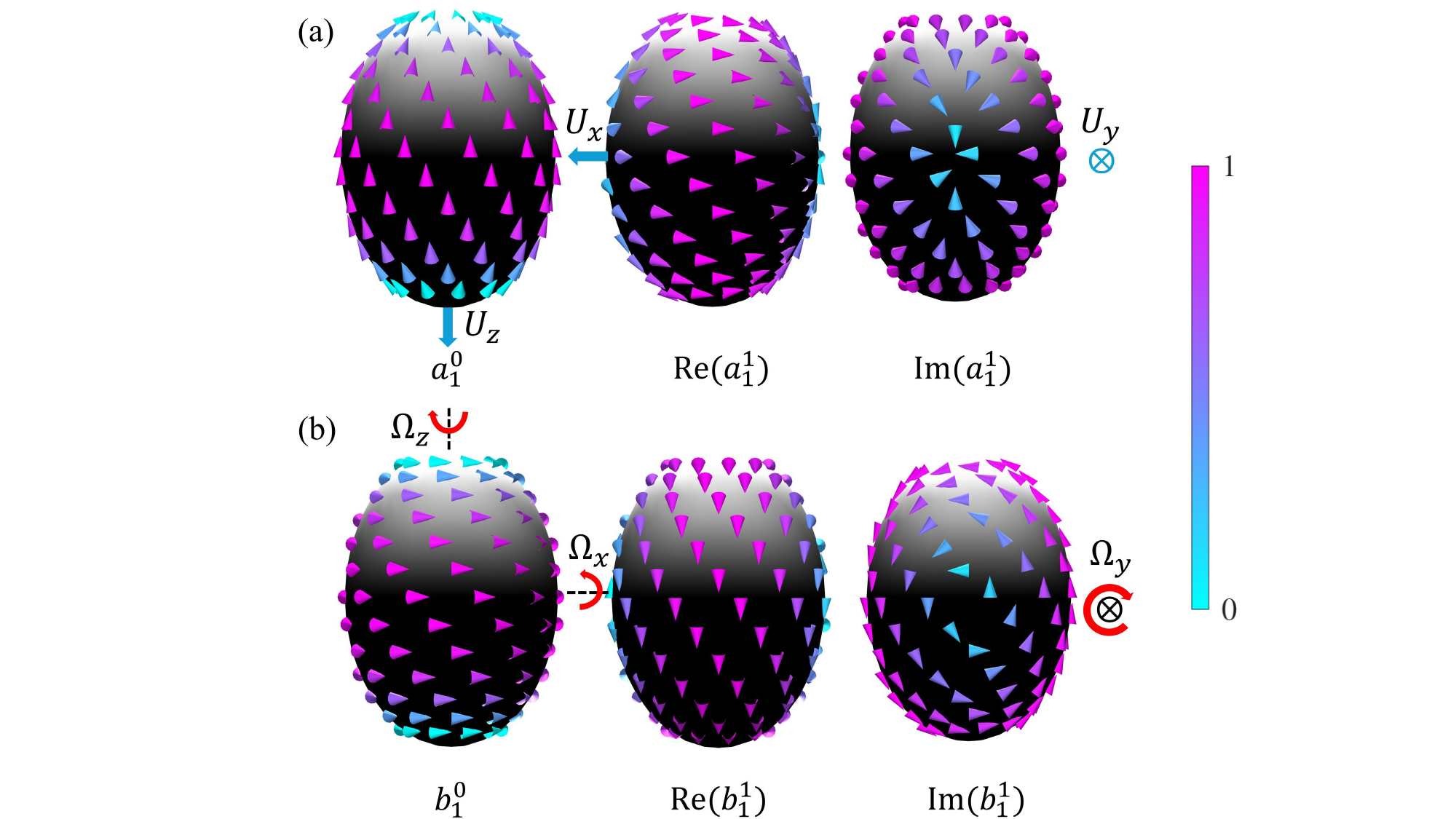}}
    \caption{Summary of first-order modal behavior. Slip velocity profiles and resultant rigid body velocities when all modes are set to zero except (a) $a_1^0 = 1$ (left), $a_1^1 = 1 = -a_1^{-1}$ (center), and $a_1^1 = \mathrm{i} = a_1^{-1}$ (right) and (b) $b_1^0 = 1$ (left), $b_1^1 = 1 = -b_1^{-1}$ (center), and $b_1^1 = \mathrm{i} = b_1^{-1}$ (right). Colour bar represents the magnitude of the normalised slip velocity. }
    \label{fig: firstordersummary}
\end{figure}

The above results demonstrate the decoupling of the first-order modes. Each component of the rigid body velocity is determined by exactly one of the coefficients in \eqref{eq: slipdecompfirstorder}. The $a_1^m$ modes determine the translational velocity $\boldsymbol{U}$, while the $b_1^m$ modes determine the angular velocity $\boldsymbol{\Omega}$. The $m = 0$ modes produce rigid body motion in the $z$-direction, while the $m = 1$ modes allow for translation/rotation in the $x$- and $y$-directions. For a given spheroid, the mode coefficients can be chosen to produce any desired rigid body velocity pair $(\boldsymbol{U},\boldsymbol{\Omega})$.  A summary of this first-order modal behavior is shown in figure \ref{fig: firstordersummary}. An analogous decoupling of modes is observed for spherical chiral squirmers \citep{Buradachiral,samatas2023hydrodynamic}. 
In fact, taking the limit as $\xi \rightarrow 1^{+}$, the prolate spheroid approaches a sphere of radius $a$ and the rigid body velocities take the form 
\begin{align}
\lim_{\xi \to 1^+} \boldsymbol{U} &= \frac{1}{3a}\sqrt{\frac{3}{\pi}} \Big( \sqrt{2}\mathrm{Re}(a_1^1), \, -\sqrt{2}\mathrm{Im}(a_1^1), \, -a_1^0 \Big), \\
\lim_{\xi \to 1^+} \boldsymbol{\Omega} &= \frac{1}{2a^2}\sqrt{\frac{3}{\pi}} \Big( \sqrt{2}\mathrm{Re}(b_1^1), \, -\sqrt{2}\mathrm{Im}(b_1^1), \, -b_1^0 \Big),
\end{align}
in agreement with the results for spherical chiral squirmers in \cite{Buradachiral}. 

\begin{figure}
    \centering
    \includegraphics[trim=200 0 200 0, clip, width=0.8\textwidth]{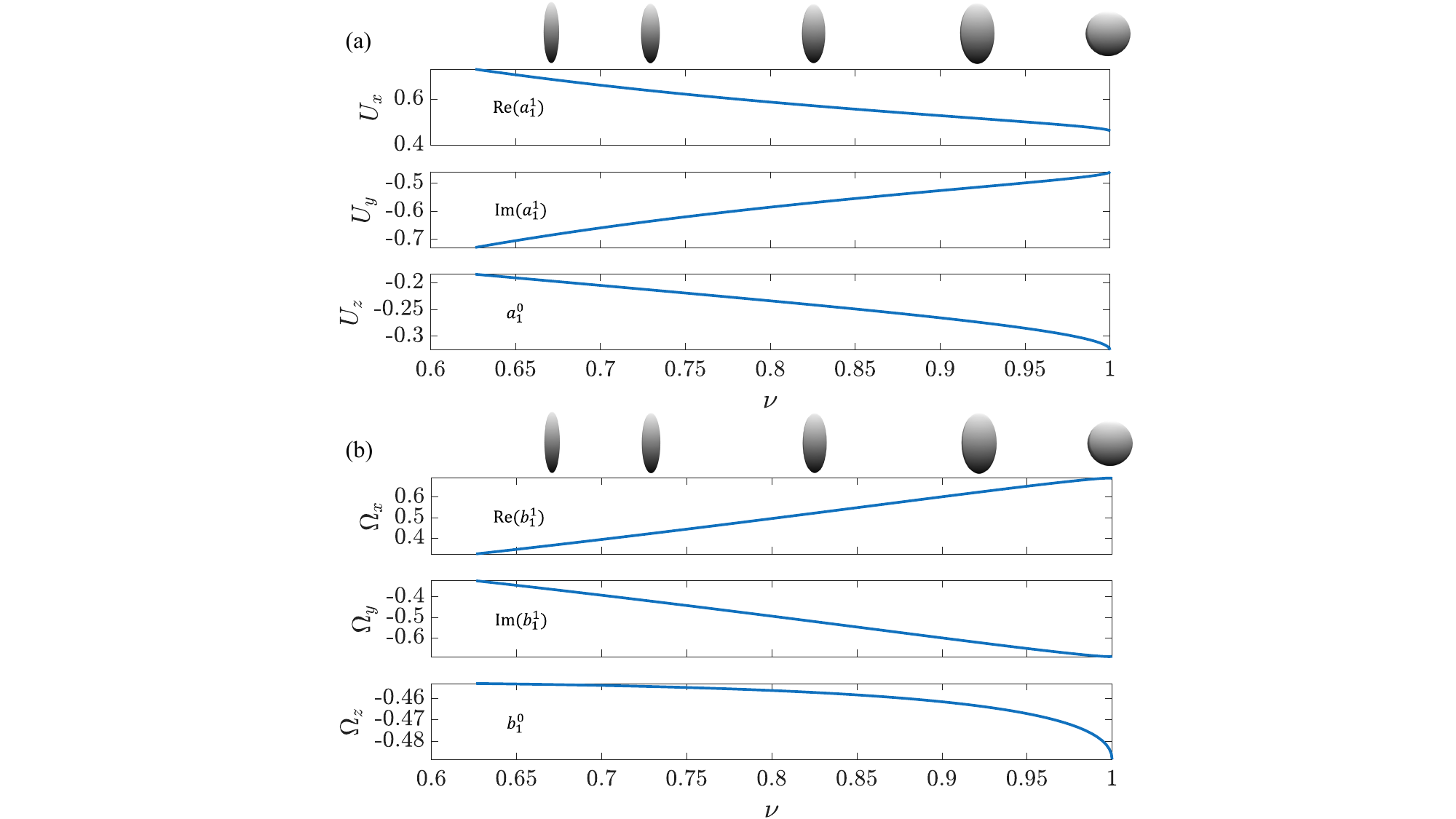}
    \caption{Shape dependence of rigid body velocities due to each mode. (a) Plots of $U_x$, $U_y$, and $U_z$ versus reduced volume $\nu$ when all modes are set to zero except $a_1^1 = 1 = -a_1^{-1}$ (top), $a_1^1 = \mathrm{i} = a_1^{-1}$ (middle), and $a_1^0 = 1$ (bottom), respectively. (b) Plots of $\Omega_x$, $\Omega_y$, and $\Omega_z$ versus reduced volume $\nu$ when all modes are set to zero except $b_1^1 = 1 = -b_1^{-1}$ (top), $b_1^1 = \mathrm{i} = b_1^{-1}$ (middle), and $b_1^0 = 1$ (bottom). Spheroids have constant surface area $4 \pi$. As $\nu$ decreases, $U_z$, $\Omega_x,$ $\Omega_y$, $\Omega_z$ all decrease in magnitude, while $U_x$ and $U_y$ increase in magnitude.}
    \label{fig: reducedvolrbv}
\end{figure}

More generally, we can observe how the rigid body velocities vary as the aspect ratio $\xi$ is varied. In figure \ref{fig: reducedvolrbv}, each squirming mode is activated individually and the resulting rigid body velocity component is plotted as a function of the reduced volume $\nu = 6 \sqrt{\pi} V/ A^{3/2}$, where $V$ and $A$ are the volume and surface area of the spheroid, respectively. As the aspect ratio increases, the spheroid becomes more elongated along the $z$-axis and the reduced volume decreases. As $\nu$ decreases, $U_z$, $\Omega_x,$ $\Omega_y$, $\Omega_z$ all decrease in magnitude, while $U_x$ and $U_y$ increase in magnitude, suggesting that a higher aspect ratio leads to reduced rotation in all directions and translation in the $z$-direction, and enhanced translation in the $x$- and $y$-directions. 

\section{Slip optimization}
\label{sc:optim}
 The prolate spheroid case study of the previous section provides an improved understanding of how the individual squirming modes affect the motion of the swimmer. In this section, we aim to answer the inverse problem: ``{\em Given a microswimmer of arbitrary shape $\Gamma$, what slip velocity minimizes the power loss?}''

We define the power loss functional 
\begin{align}
P(\boldsymbol{v}^{\textrm{S}}) := \langle \boldsymbol{f}, \boldsymbol{v}^{\textrm{S}} + \boldsymbol{v}^{\textrm{R}} \rangle_\Gamma = \langle \boldsymbol{f}, \boldsymbol{v}^{\textrm{S}}\rangle_\Gamma,
\end{align}
where the second equality is due to the force-free and torque-free conditions \eqref{eq: forcetorquefree}. 

\begin{figure}
    \centering
    \includegraphics[trim=60 40 40 0, clip, width=0.8\textwidth]{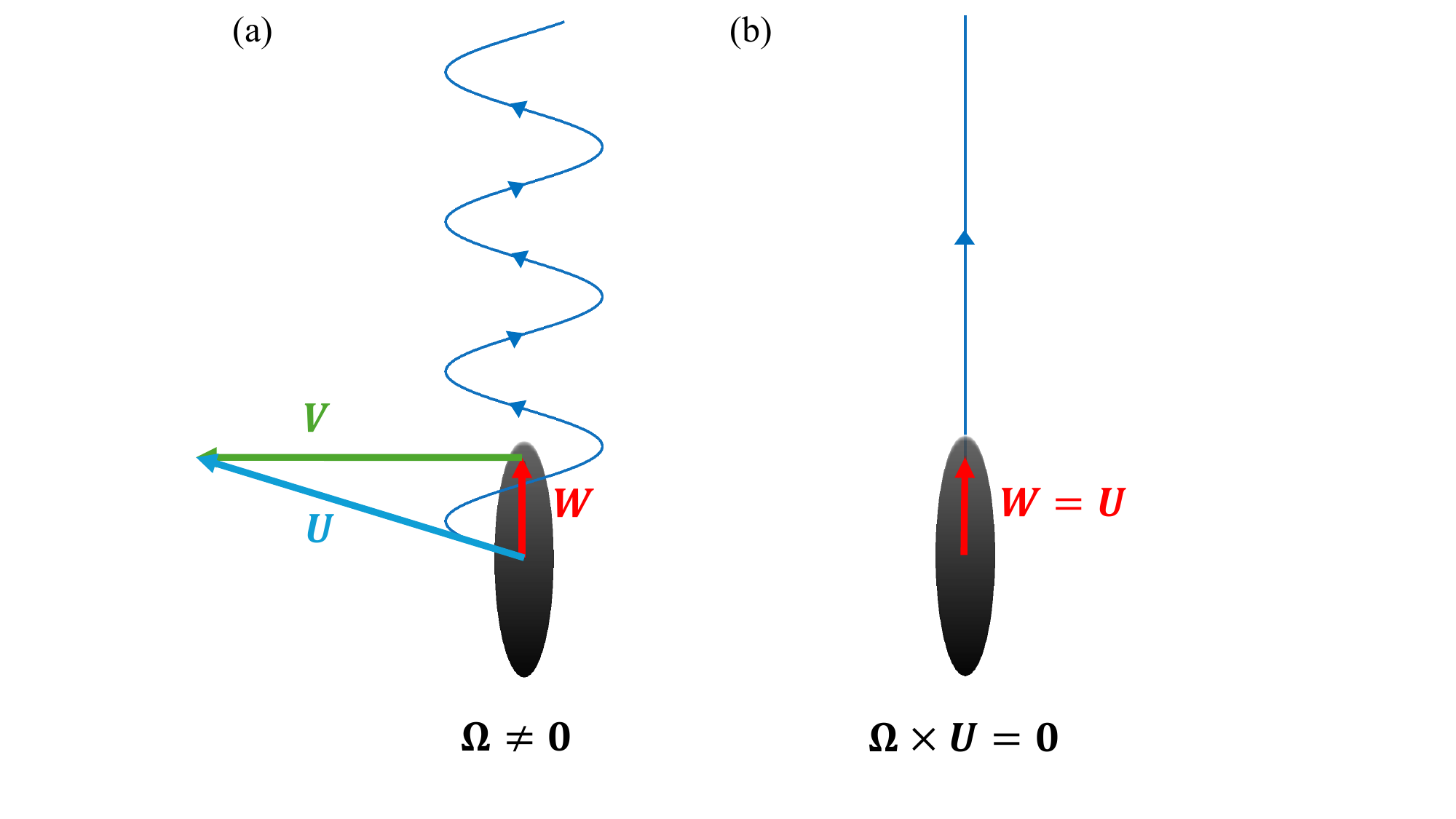}
    \caption{Plots of swimmer motion as well as the translational velocity $\boldsymbol{U}$ and its components $\boldsymbol{V}$ and $\boldsymbol{W}$ for (a) helical motion with $\boldsymbol{\Omega} \neq \boldsymbol{0}$ and $\boldsymbol{V} \neq \boldsymbol{0}$  and (b) straight line motion with $\boldsymbol{\Omega} \times \boldsymbol{U} = \boldsymbol{0}$ and $\boldsymbol{V} = \boldsymbol{0}$.} 
    \label{fig:caseab}
\end{figure}

In the axisymmetric case, the power loss can be made arbitrarily small by reducing $|\boldsymbol{U}|$, so the swimming speed $|\boldsymbol{U}|$ is fixed \citep{guo2021optimal}. Since arbitrary shape swimmers can rotate, the angular velocity $\boldsymbol{\Omega}$ provides extra degrees of freedom. Moreover, as seen in Section \ref{sc:helix},  $\boldsymbol{\Omega}$ determines the direction of the helix axis, i.e. the direction of net motion for the swimmer. In this vein, rather than fixing $|\boldsymbol{U}|$, we fix $|\boldsymbol{W}|$, where $\boldsymbol{W}$ is the translational velocity \textit{along the helix axis (direction of net motion)}. In biological settings, $\boldsymbol{W}$ may represent the direction of an external stimulus. Just as we have seen in Section \ref{sc:helix}, $\boldsymbol{W}$ takes a different form depending on the values of $\boldsymbol{U}$ and $\boldsymbol{\Omega}$:
\begin{equation}
   \boldsymbol{W} = \frac{\boldsymbol{U} \boldsymbol{\cdot} \boldsymbol{\Omega}}{|\boldsymbol{\Omega}|^2} \boldsymbol{\Omega} \quad(\textrm{if } \boldsymbol{\Omega} \neq \boldsymbol{0}), \qquad 
  \boldsymbol{W} = \boldsymbol{U} \quad(\textrm{if }\boldsymbol{\Omega} \times \boldsymbol{U} = \boldsymbol{0}). 
\label{eq: W}
\end{equation}
The first case covers all helical and rotating straight line motions, while the second covers all straight line motions (rotating and non-rotating) (see figure \ref{fig:caseab}). Fixing $|\boldsymbol{W}| = 1$ (unit speed along the helix axis), the optimal slip velocity that minimizes power loss solves the constrained minimization problem 
\begin{align}
\min_{\boldsymbol{v}^\textrm{S}\in H_T}P(\boldsymbol{v}^\textrm{S}) \quad \textrm{subject to} \quad|\boldsymbol{W}| = 1
\label{eq: constrmin}
\end{align}
where $H_T$ is the space of all tangential vector fields on $\Gamma$.

In fact, the companion paper by~\cite{B-2025-06} shows that the finite-dimensional nature of the induced rigid-body motions allows one to reduce the search space $H_T$ of problem~\eqref{eq: constrmin} to a specific 6-dimensional subspace $H_6$ of tangential vector fields, whose basis functions depend only on the swimmer shape and are defined from the solutions of an additional set of six auxiliary Stokes flow problems given therein. For arbitrary shapes, these solutions are obtained using a spectrally-accurate solver based on boundary integral equations \citep{veerapaneni2011fast, rahimian2015boundary}. 

In addition, the rigid-body motion parameters are reparametrized by observing that the translational velocity $\boldsymbol{U}$ can be decomposed into its projection along the helix axis $\boldsymbol{W}$ and a component $\boldsymbol{V}$ that is orthogonal to $\boldsymbol{W},$ i.e. $\boldsymbol{U} = \boldsymbol{W}+\boldsymbol{V}$ with $\boldsymbol{W} \boldsymbol{\cdot} \boldsymbol{V} = 0$ (see figure \ref{fig:caseab}), while $\boldsymbol{\Omega} = s \boldsymbol{W}$ for some $s \in \mathbb{R}$ since $\boldsymbol{\Omega}$ points in the direction of net motion. As a result, problem~\eqref{eq: constrmin} is recast without loss of generality in the form
\begin{align}
\min_{(s,\boldsymbol{V},\boldsymbol{W})}\mathcal{P}(s,\boldsymbol{V},\boldsymbol{W}) \quad \textrm{subject to} \quad\boldsymbol{W} \boldsymbol{\cdot} \boldsymbol{V} = 0, \quad |\boldsymbol{W}| = 1
\label{eq: optimizationsVW}
\end{align}
where $\mathcal{P}(s,\boldsymbol{V},\boldsymbol{W}) := P(\boldsymbol{v}^{\textrm{S}})$ with $\boldsymbol{v}^{\textrm{S}}\in H_6$ producing $(\boldsymbol{U},\boldsymbol{\Omega})$ through~\eqref{eq: matrixeqn}. Moreover, $(s,\boldsymbol{V})\mapsto\mathcal{P}(s,\boldsymbol{V},\boldsymbol{W})$ is found to be quadratic (while $(s,\boldsymbol{V},\boldsymbol{W})\mapsto\mathcal{P}(s,\boldsymbol{V},\boldsymbol{W})$ is only homogeneous with degree 2), an observation that leads one to treat problem~\eqref{eq: optimizationsVW}
as two nested minimization problems. We first perform a \textit{partial optimization} in which power loss is minimized when the net motion direction $\boldsymbol{W}$ is prescribed:
\begin{align}
\min_{(s,\boldsymbol{V})}\mathcal{P}(s,\boldsymbol{V},\boldsymbol{W}) \quad \textrm{subject to} \quad\boldsymbol{W} \boldsymbol{\cdot} \boldsymbol{V} = 0.
\label{eq: optimizationsV}
\end{align}
For each $\boldsymbol{W}$ there is a corresponding minimizer $(s[\boldsymbol{W}],V[\boldsymbol{W}])$ for \eqref{eq: optimizationsV}. 
Being quadratic with a linear constraint, problem \eqref{eq: optimizationsV} has a closed-form solution $(s[\boldsymbol{W}],V[\boldsymbol{W}])$, see details in~\citet{B-2025-06}. Then, letting $\widehat{\mathcal{P}}(\boldsymbol{W}) := \mathcal{P}(s[\boldsymbol{W}],V[\boldsymbol{W}],\boldsymbol{W})$ be the optimal power loss for the partial minimization problem \eqref{eq: optimizationsV}, problem~\eqref{eq: optimizationsVW} further reduces to a \textit{global optimization} problem in which the optimal net motion direction $\boldsymbol{W}_{\textrm{opt}}$ is determined by solving
\begin{align}
\min_{\boldsymbol{W}}  \widehat{\mathcal{P}}(\boldsymbol{W}) \quad \textrm{subject to}  \quad |\boldsymbol{W}| = 1.
\label{eq: optimizationW}
\end{align}
The objective function $\widehat{\mathcal{P}}$ is homogeneous with degree 2 but in general not quadratic. Hence, problem \eqref{eq: optimizationW} does not have an obvious closed form solution, but a numerical solution can be obtained using standard optimization toolboxes.

For the purposes of this paper, the proofs and solution algorithm for the partial and global optimization problems are omitted -- we refer the reader to~\cite{B-2025-06} for the details.

\section{Results and discussion}
\label{sc:results}

Next, we present the results of the slip optimization procedure for various shapes. Any shape parametrizations that are not explicitly given in this section can be found in the shape gallery in Appendix \ref{ap: shape gallery}. We use the minimal power loss $12 \pi R_0$ of a spherical squirmer of radius $R_0 = \sqrt{|\Gamma|/(4 \pi)}$ moving at unit speed to normalize all the power loss values.

The reflectional symmetries of the swimmer shape $\Gamma$
play a key role in determining whether the optimal motion is purely translational or involves rotation. Let us first test axisymmetric shapes by setting $\boldsymbol{W} = \boldsymbol{e}_z$ so that the swimmer moves along its axis of symmetry. The partial optimization yields a non-rotating straight line solution with $\boldsymbol{U} = \boldsymbol{W}$ and $\boldsymbol{\Omega} = \boldsymbol{0}$, as seen in figure \ref{fig:axiopt}(a) for the axisymmetric stomatocyte (see Appendix \ref{ap: shape gallery} for the shape equation). This agrees with the observation in \cite{stone1996propulsion} that an axisymmetric swimmer that rotates as it swims is less efficient than the non-rotating swimmer. The optimal slip profiles found match the ones in \cite{guo2021optimal} for axisymmetric slip optimization (see figure \ref{fig:axiopt}(b)).

\begin{figure}
    \centering
    \includegraphics[trim=10 140 0 30, clip, width=0.9\textwidth]{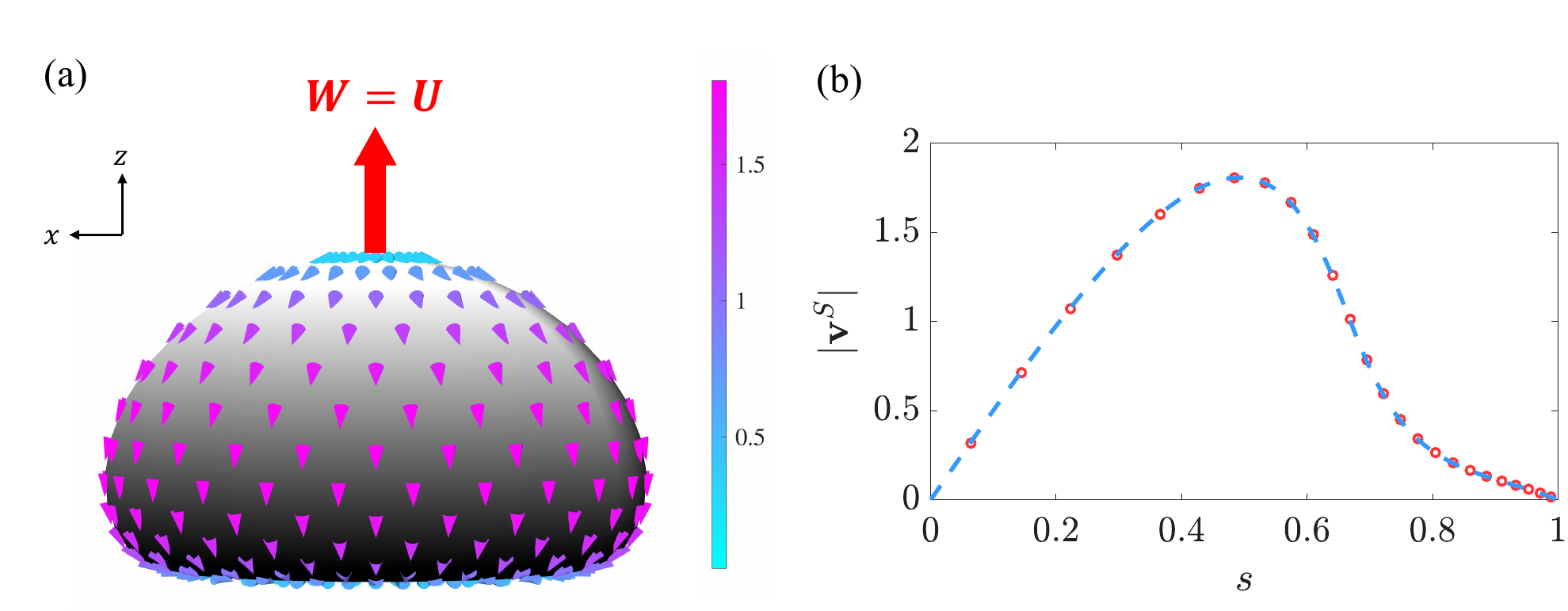}
    \caption{Partial slip optimization of the axisymmetric stomatocyte (with $\lambda = 0.4$) when $\boldsymbol{W} = \boldsymbol{e}_z$ points along its axis of symmetry. (a) Optimal slip velocity field plotted together with the direction of the net motion $\boldsymbol{W} = \boldsymbol{U}$ (red arrow). The angular velocity is computed to be $\boldsymbol{\Omega} = \boldsymbol{0}$, resulting in rotation-free, straight line motion along $\boldsymbol{W}$. The colour bar represents $|\boldsymbol{v}^\textrm{S}|$. (b) Plot of $|\boldsymbol{v}^\textrm{S}|$ as a function of normalised arc length of the generating curve $s$. The arbitrary shape optimizer results (red circles) agree with the axisymmetric optimizer results of \cite{guo2021optimal} (blue dotted line).} 
    \label{fig:axiopt}
\end{figure}

 More generally, we can study arbitrary shapes by considering the family of shapes parametrized by 
\begin{align}
\boldsymbol{x}(\theta,\phi) = \rho(\theta,\phi)(\sin \theta \cos \phi, \sin \theta \sin \phi, \cos \theta).
\label{family}
\end{align}
The function $\rho(\theta,\phi)$ in \eqref{family} is of particular importance as it can be used to deduce if the $z = 0$, $x = 0$, and $y = 0$ planes are reflectional planes of symmetry for a shape:
\begin{itemize}
\item If $\rho(\pi-\theta,\phi) = \rho(\theta,\phi)$, then the $z=0$ plane is a plane of symmetry,
\item If $\rho(\theta, \pi-\phi) =  \rho(\theta,\phi)$,   then the $x=0$ plane is a plane of symmetry,
\item If $\rho(\theta, -\phi) =  \rho(\theta,\phi)$,  then the $y = 0$ plane is a plane of symmetry.  
\end{itemize}
Clearly, shapes may have planes of symmetry that are not one of these three planes. Such other symmetries may be described similarly. For instance, $\rho(\theta,\pi+\phi) = \rho(\theta,\phi)$ implies central symmetry in the $(x,y)$ plane (half-turn about the $z$-axis), and more generally $\rho(\theta,\phi+2\pi/m) = \rho(\theta,\phi)$ implies cyclic symmetry of order $m$ about the $z$-axis. 

\begin{figure}
    \centering
    \includegraphics[trim=0 260 60 0, clip, width=\textwidth]{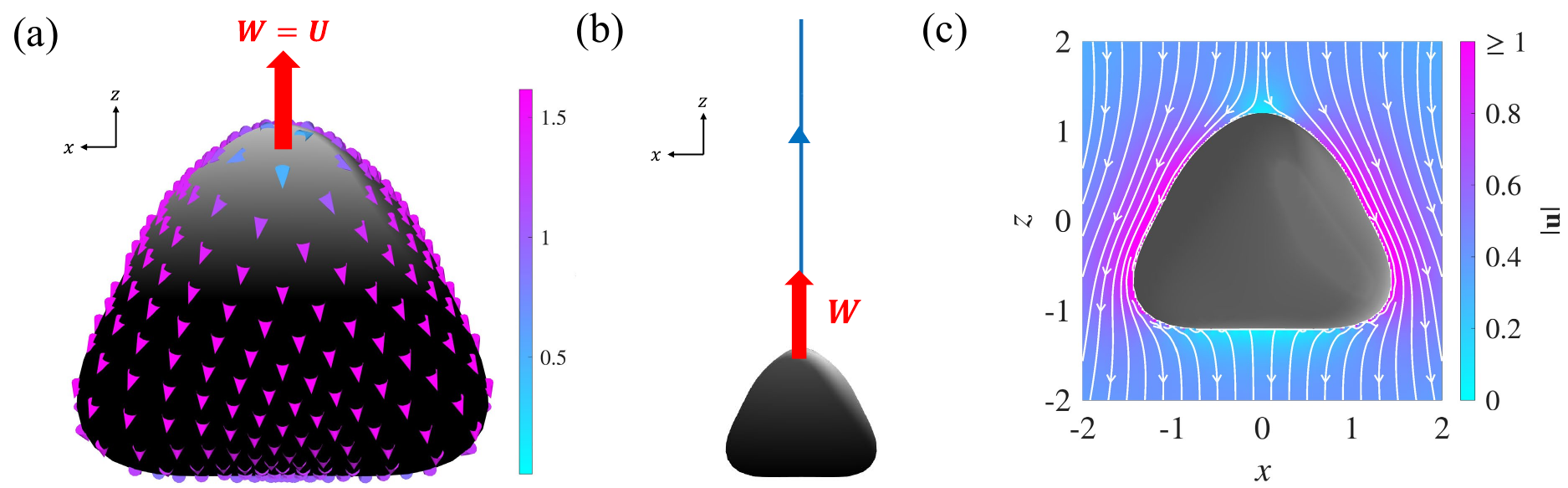}
    \caption{Partial optimization when $\boldsymbol{W}$ lies within a symmetry plane. (a) Optimal slip velocity field for a non-axisymmetric triangular antiprism, when $\boldsymbol{W} = (0,0.7069,0.7073)$ (red arrow) lies within its $x=0$ symmetry plane. The translational velocity is $\boldsymbol{U} = \boldsymbol{W} $ and the angular velocity is $\boldsymbol{\Omega} = \boldsymbol{0}$, resulting in rotation-free, straight line motion. Colour bar represents $|\boldsymbol{v}^\textrm{S}|$. (b) Resulting straight line motion in the direction of $\boldsymbol{W}$. Swimmer has been enlarged for illustrative purposes. (c) Body frame flow field streamlines in the $y = 0$ plane. The optimization resulted in a normalised minimum power loss of $\widehat{\mathcal{P}}(\boldsymbol{W})/(12 \pi R_0) = 1.1267$. }
    \label{fig: inplaneopt}
\end{figure}

Based on the above classification, the non-axisymmetric triangular antiprism (see Appendix \ref{ap: shape gallery}) has the $y=0$ and $x = 0$ planes as planes of symmetry. When $\boldsymbol{W}$ lies in one of these planes, the partial optimization once again gives a solution with   $\boldsymbol{U} = \boldsymbol{W}$ and $\boldsymbol{\Omega} = \boldsymbol{0}$ (see figure \ref{fig: inplaneopt}). More generally, a non-rotating optimal solution arises when  $\boldsymbol{W}$ lies in a symmetry plane of the shape. Since the swimmer is achiral with respect to this symmetry plane, the optimal motion is non-rotating straight line motion as it respects the reflectional symmetry. A mathematical justification for this observation is given in~\cite{B-2025-06}.


\begin{figure}
    \centering
    \includegraphics[trim=0 0 0 0, clip, width=\textwidth]{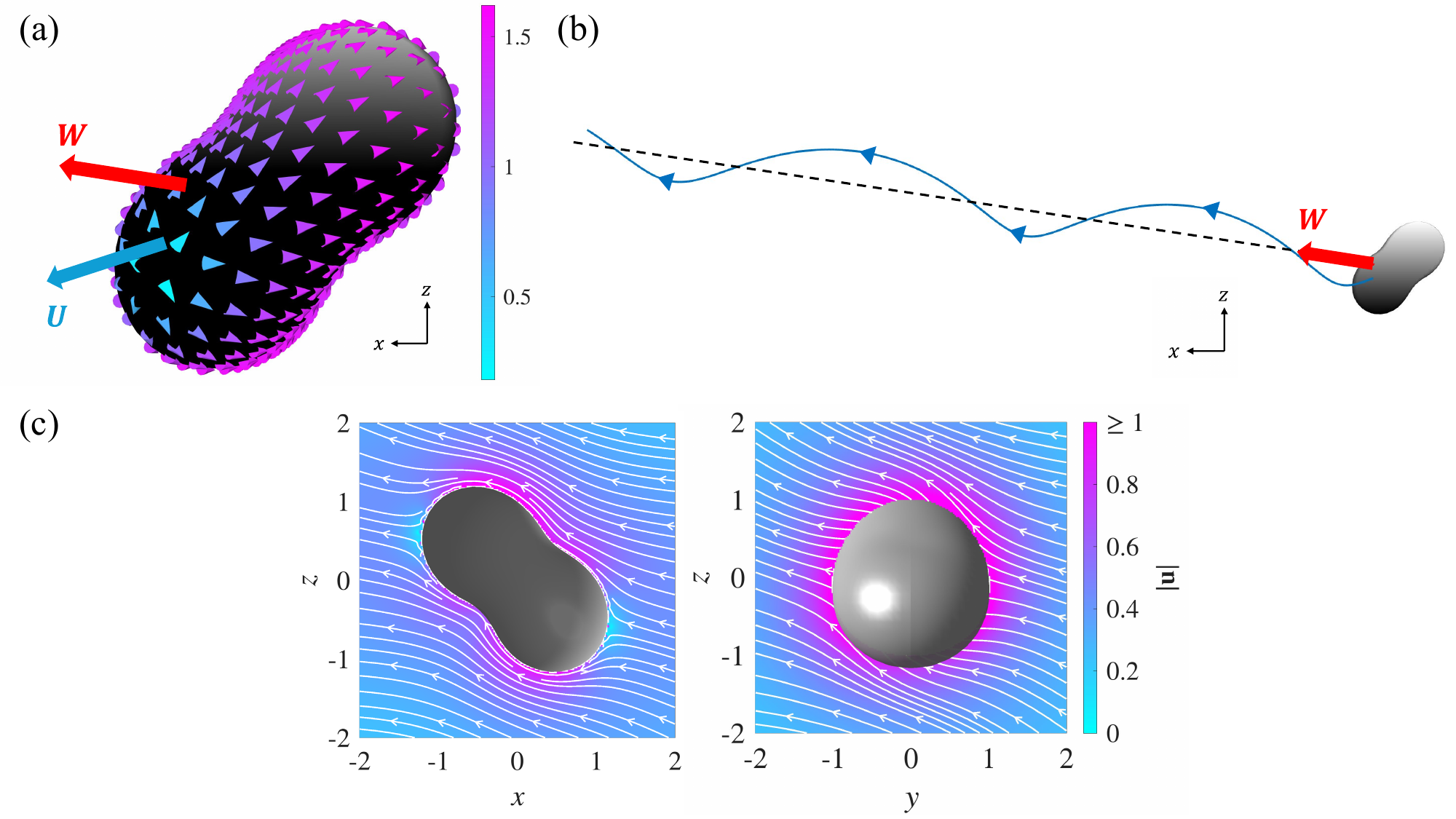}
    \caption{Partial optimization when $\boldsymbol{W}$ lies outside of a symmetry plane. (a) Optimal slip velocity field for a tilted dumbbell when $\boldsymbol{W} = (0.6651,0.7395,0.1037) =\boldsymbol{\Omega}/|\boldsymbol{\Omega}|$ (red arrow) does not lie in a symmetry plane. The translational velocity is $\boldsymbol{U} = (0.7042, 0.7535, -0.2468)$ (blue arrow). The nonzero angular velocity $\boldsymbol{\Omega} = (0.0353,0.0392,0.0055)$ results in helical motion. Colour bar represents $|\boldsymbol{v}^\textrm{S}|$. (b) Resulting helical motion about the helical axis $\boldsymbol{W}$. Swimmer has been enlarged for illustrative purposes. (c) Body frame flow field streamlines in the $y = 0$ (left) and $x = 0$ (right) planes. The optimization resulted in a normalised minimum power loss of $\widehat{\mathcal{P}}(\boldsymbol{W})/(12 \pi R_0) =1.0659$. }
    \label{fig: tilt_dumbbell_partialopt}
\end{figure}

When $\boldsymbol{W}$ lies outside of a symmetry plane, 
the partial minimization~\eqref{eq: optimizationsV}
can produce nonzero $\boldsymbol{\Omega}$. The tilted dumbbell shape has the $y = 0$ plane as a symmetry plane, and choosing $\boldsymbol{W}$ to lie outside of this plane leads to an optimal motion that is helical (see figure \ref{fig: tilt_dumbbell_partialopt}). {\em Thus, unlike in the axisymmetric case, rotational motion can help reduce the total power loss for certain directions of net motion.}
 
\begin{figure}
    \centering
    \includegraphics[trim=0 130 0 80, clip, width=\textwidth]{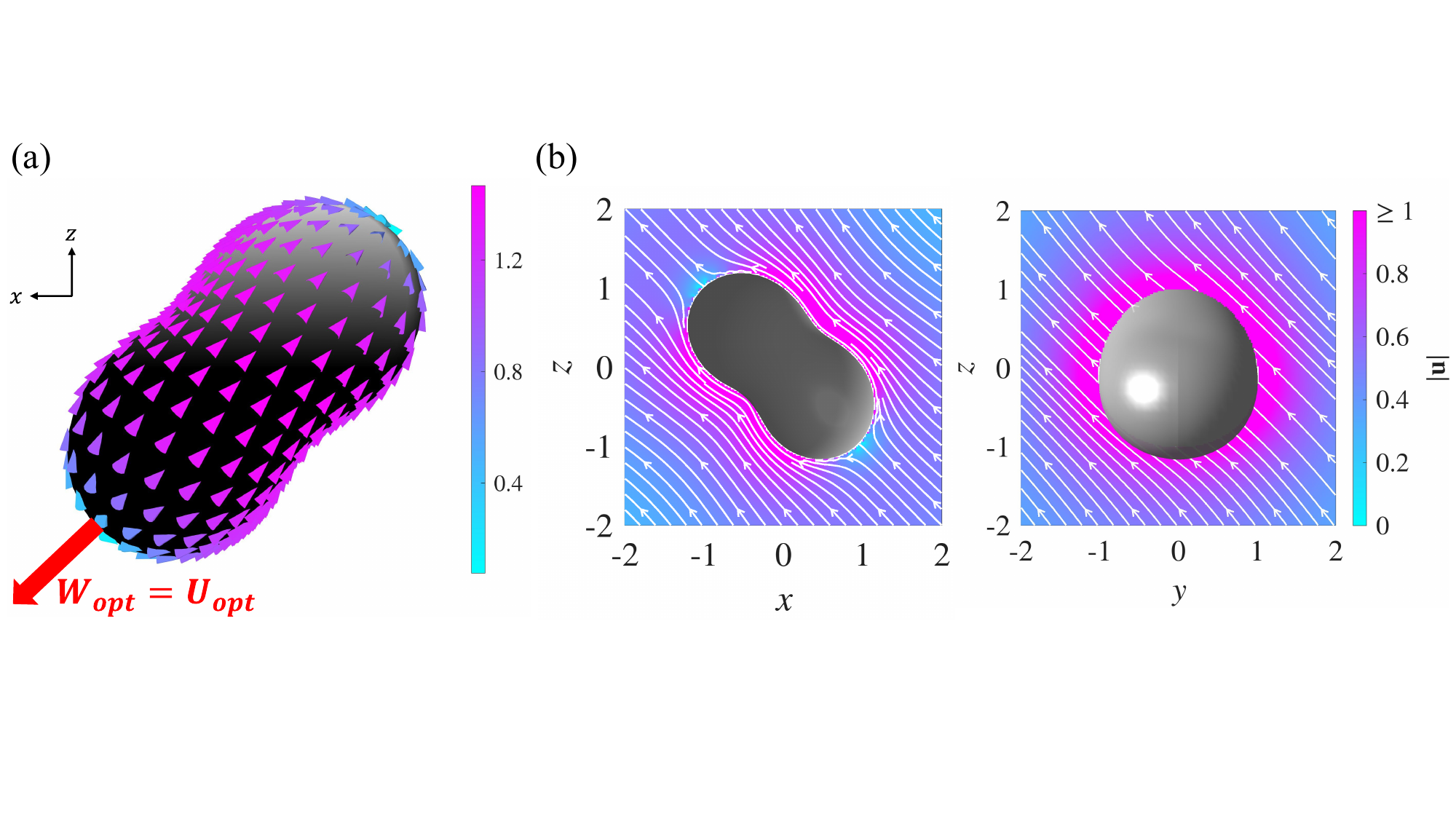}
    \caption{Global optimization of a tilted dumbbell. (a) Optimal slip velocity field plotted together with its corresponding optimal direction of net motion $\boldsymbol{W}_{\textrm{opt}} =  \boldsymbol{U}_{\textrm{opt}} = (0.7069,0,-0.7073)$ (red arrow). The angular velocity is computed to be $\boldsymbol{\Omega}_{\textrm{opt}} = \boldsymbol{0}$, resulting in rotation-free, straight line motion along $\boldsymbol{W}_\textrm{opt}$. Colour bar represents $|\boldsymbol{v}^\textrm{S}|$. (b) Body frame flow field streamlines in the $y = 0$ (left) and $x = 0$ (right) planes. The optimization resulted in a normalised global minimum power loss of $\widehat{\mathcal{P}}(\boldsymbol{W}_\textrm{opt})/(12 \pi R_0) = 0.6304$, a $41 \%$ reduction from figure \ref{fig: tilt_dumbbell_partialopt}.}
    \label{fig: tilt_dumbbell_globalopt}
\end{figure}

However, global optimization~\eqref{eq: optimizationW} for the tilted dumbbell gives a globally optimal direction of net motion $\boldsymbol{W}_{\textrm{opt}}$ that lies within its $y = 0$ symmetry plane, leading to achiral, straight line motion (see figure \ref{fig: tilt_dumbbell_globalopt}). In this case, the globally optimal motion respects the symmetry of the swimmer, and leads to a nearly $41 \%$ reduction in power loss compared to the partial optimization example in figure \ref{fig: tilt_dumbbell_partialopt}.

\begin{figure}
    \centering
    \includegraphics[trim=0 170 80 0, clip, width=0.8\textwidth]{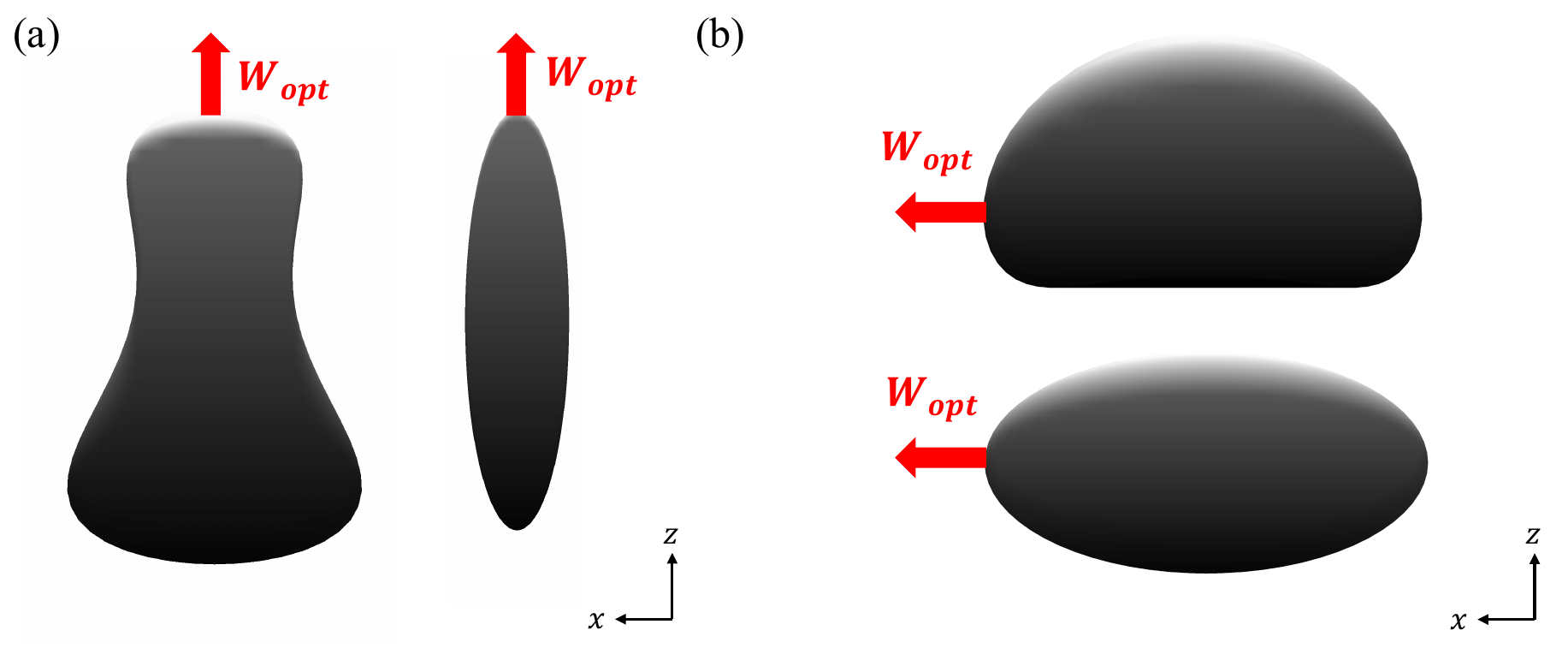}
    \caption{Global optimization for axisymmetric shapes. For all the axisymmetric shapes, the optimal net motion directions satisfy $\boldsymbol{W}_{\textrm{opt}} = \boldsymbol{U}_{\textrm{opt}}$ (red arrows) and corresponding angular velocity is $\boldsymbol{\Omega}_{\textrm{opt}}=\boldsymbol{0}$ (non-rotating straight line motion). (a) For prolate shapes that are slender along the $z$-axis, like the butternut (left) and prolate spheroid (right) we have $\boldsymbol{W}_{\textrm{opt}} = \boldsymbol{e}_z$. (b) For oblate shapes, like the stomatocyte (top) and oblate spheroid (bottom), $\boldsymbol{W}_{\textrm{opt}}$ can be any vector in the $z = 0$ plane. }
    \label{fig: axiglobalopt}
\end{figure}

Based on the global optimization for the tilted dumbbell, one may think that the globally optimal motion respects the symmetry of the swimmer shape. The global optimization for axisymmetric shapes is an interesting case study as for an axisymmetric swimmer, every plane passing through the origin that is orthogonal to the $z = 0$ plane is a symmetry plane. Since any $\boldsymbol{W} \in \mathbb{R}^3$ will lie in one of these planes, all choices of $\boldsymbol{W}$ respect plane symmetry. Nevertheless, our intuition suggests that $\boldsymbol{W}_{\textrm{opt}}$ should lie along the rotational symmetry axis as the shape is ``most symmetric'' with respect to this axis. Indeed, our results show that for prolate shapes that are slender along the rotational axis (butternut and prolate spheroid), $\boldsymbol{W}_{\textrm{opt}}$ points along the rotational axis, as expected (figure \ref{fig: axiglobalopt}(a)). However, contrary to intuition, $\boldsymbol{W}_{\textrm{opt}}$ lies in the $z = 0$ plane, rather than the rotational axis, for the oblate shapes (stomatocyte and oblate spheroid)(see figure \ref{fig: axiglobalopt}(b)). In both the prolate and oblate cases, since $\boldsymbol{W}_{\textrm{opt}}$ always lies in a symmetry plane, the globally optimal motion solving~\eqref{eq: optimizationW} is always a non-rotating straight line with $\boldsymbol{W}_{\textrm{opt}} =\boldsymbol{U}_{\textrm{opt}} $ and $\boldsymbol{\Omega}_{\textrm{opt}} = \boldsymbol{0}$. To determine whether $\boldsymbol{W}_{\textrm{opt}}$ lies along the $z$-axis or in the $z=0$ plane, shape-dependent constants can be computed and compared~\cite{B-2025-06}.

\begin{figure}
    \centering
    \includegraphics[trim=0 0 10 0, clip, width=\textwidth]{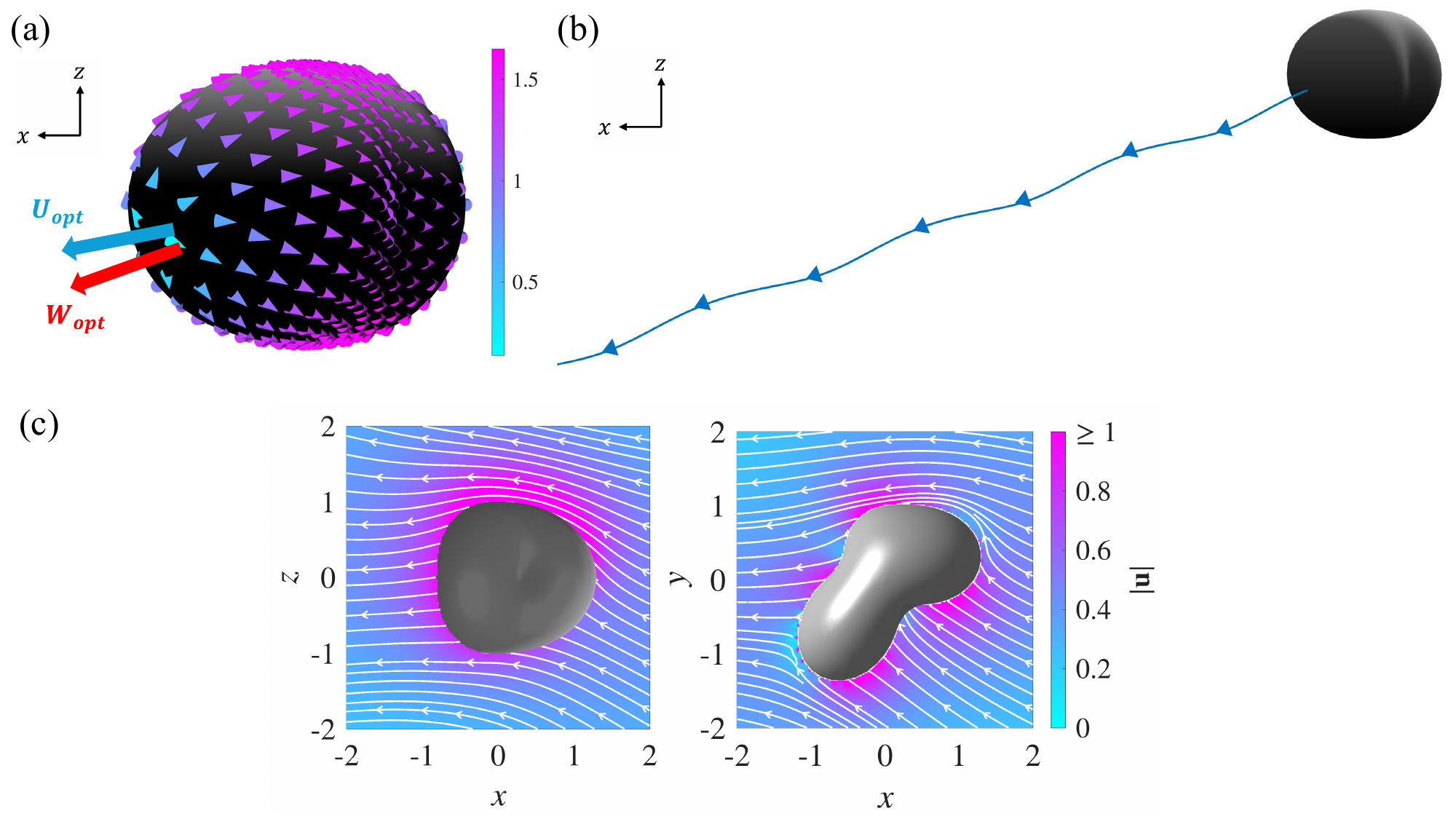}
    \caption{Global optimization of the shape described by \eqref{eq: skewboomerang}. (a) Optimal slip velocity field plotted together with its corresponding optimal direction of net motion $\boldsymbol{W}_{\textrm{opt}} = \boldsymbol{\Omega_\textrm{opt}}/|\boldsymbol{\Omega_\textrm{opt}}| =  (0.7071,0.6616,-0.2496) $ (red arrow) and the direction of the translational velocity $\boldsymbol{U_\textrm{opt}} = (0.7343,0.6749,-0.1373)$ (blue arrow). The nonzero angular velocity $\boldsymbol{\Omega_\textrm{opt}} = (0.3544,0.3316, -0.1251)$ results in globally optimal helical motion. Colour bar represents $|\boldsymbol{v}^\textrm{S}|.$ (b) Resulting helical motion. Swimmer has been enlarged for illustrative purposes. (c) Body frame flow field streamlines in the $y = 0$ (left) and $z = 0$ (right) planes. The optimization resulted in a normalised global minimum power loss of $\widehat{\mathcal{P}}(\boldsymbol{W}_\textrm{opt})/(12 \pi R_0) = 0.7396 $.}
    \label{fig: skewboomerang_globalopt}
\end{figure}

From the previous example, we see that the globally optimal direction of net motion $\boldsymbol{W}_{\textrm{opt}}$ need not be the most symmetric choice. It is tempting to conclude that like the tilted dumbbell and axisymmetric examples, the globally optimal motion is always non-rotating, that is we always have $\boldsymbol{\Omega}_{\textrm{opt}}=\boldsymbol{0}$. However, this does not hold in general either: for example, the shape given by 
\begin{align}
\rho(\theta,\phi) = 1 + 0.4 \sin^2(\theta)\sin(2\phi) + 0.2 \sin^3(\theta) \cos(3\phi)
\label{eq: skewboomerang}
\end{align}
leads to globally optimal motion that is helical (see figure \ref{fig: skewboomerang_globalopt}) with $\boldsymbol{\Omega}_{\textrm{opt}} \neq \boldsymbol{0}$. Thus, for certain swimmer shapes, it can be beneficial to rotate, even when a direction of net motion is not specified beforehand. Note that even though the $z = 0$ plane is a plane of symmetry of \eqref{eq: skewboomerang}, $\boldsymbol{W}_{\textrm{opt}}$ does not lie in this plane, demonstrating that $\boldsymbol{W}_{\textrm{opt}}$ does not have to lie in a plane of symmetry of a shape.

The shape given by \eqref{eq: skewboomerang} provides a prototypical example of globally optimal rotational motion. To further study how the rotational motion arises, we generalise to the family of shapes given by 
\begin{align}
\rho(\theta,\phi) = N(\alpha, \beta)(1 + \alpha \sin^2(\theta)\sin(2\phi) + \beta \sin^3(\theta) \cos(3\phi)),
\label{eq: skewboomerangfamily}
\end{align}
where $N(\alpha, \beta)$ is a normalizing factor that ensures that the surface area is $|\Gamma|=4 \pi$. Just as in \eqref{eq: skewboomerang}, all shapes in this family are symmetric with respect to $z=0$.

To better understand the global optimization problem, we plot the solution to the partial optimization problem $\widehat{\mathcal{P}}(\boldsymbol{W})$ as a function of $\boldsymbol{W}$ for various shapes in the family. The global minimum of $\widehat{\mathcal{P}}(\boldsymbol{W})$ subject to $|\boldsymbol{W}| = 1$ gives the solution to the global optimization problem -- since $\boldsymbol{W}$ lies on the unit sphere, we use spherical coordinates $(\theta,\phi)$ and write $\boldsymbol{W} = (\sin \theta \cos \phi, \sin \theta \sin \phi, \cos \theta)$. Furthermore, since $\widehat{\mathcal{P}}(\boldsymbol{W}) = \widehat{\mathcal{P}}(-\boldsymbol{W})$ \citep{B-2025-06}, we can further restrict  $\boldsymbol{W}$ to the half-sphere, i.e. $(\theta, \phi) \in [0,\pi]^2$.

\begin{figure}
    \centering
    \includegraphics[trim=60 20 80 30, clip, width=0.85\textwidth]{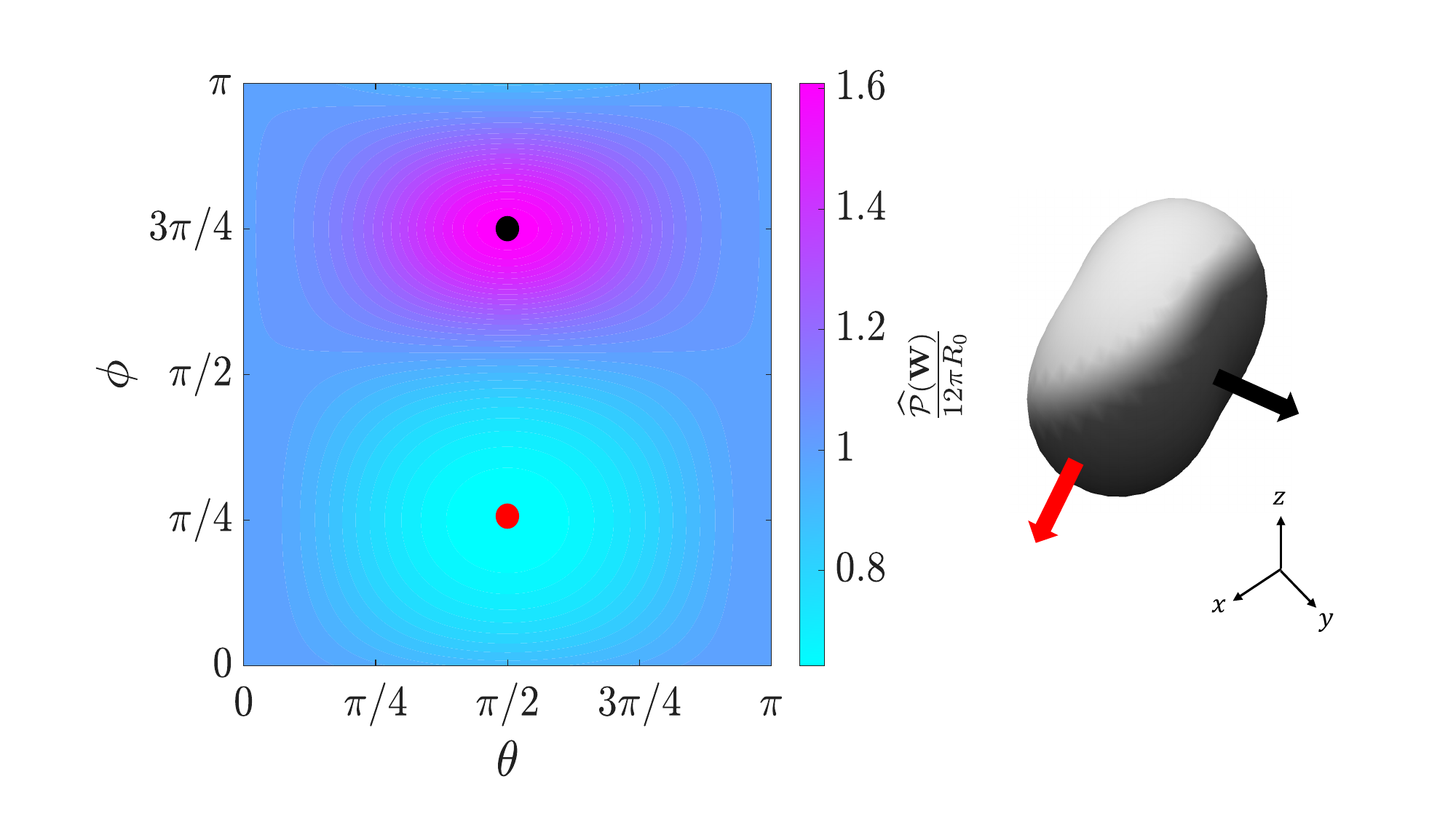}
    \caption{Contour plot of the partially minimized power loss $\widehat{\mathcal{P}}(\boldsymbol{W})$ as a function of the spherical coordinates $(\theta,\phi)$ when $\alpha = 0.3$ and $\beta = 0$. A global minimum is attained at $(\theta,\phi)\approx (\pi/2,\pi/4)$ (red point) and a global maximum is attained at $(\theta,\phi)\approx (\pi/2,3\pi/4)$ (black point). The corresponding minimizing and maximizing net motion directions are drawn as red and black arrows, respectively and are both located within the $z = 0$ symmetry plane.}
    \label{fig: twolobecontour}
\end{figure}

A contour plot of $\widehat{\mathcal{P}}(\boldsymbol{W})$ when $\alpha = 0.3$ and $\beta = 0$ is shown in figure \ref{fig: twolobecontour}. With this choice of parameters, we have a symmetric two-lobed dumbbell shape -- more precisely, in addition to its symmetry with respect to $z = 0$, it has order-2 cyclic symmetry. The power loss landscape reveals a global minimum at $(\theta,\phi)\approx (\pi/2,\pi/4)$, located in the $z=0$ symmetry plane. In the same symmetry plane, a global maximum occurs at $(\theta,\phi)\approx (\pi/2,3\pi/4)$, corresponding to moving along the midplane that separates the two lobes. This demonstrates that obeying symmetry does not guarantee an improvement in power loss -- in the same symmetry plane both a global minimum and a global maximum occur. The optimization landscape is otherwise well-behaved as no local optima that are not global optima occur. Since $\boldsymbol{W}_{\textrm{opt}}$ is located within the $z = 0$ symmetry plane, the globally optimal motion is a straight line with $\boldsymbol{\Omega}_{\textrm{opt}} = \boldsymbol{0}$, as shown in~\cite{B-2025-06}. 

\begin{figure}
    \centering
    \includegraphics[trim=0 20 90 40, clip, width=0.9\textwidth]{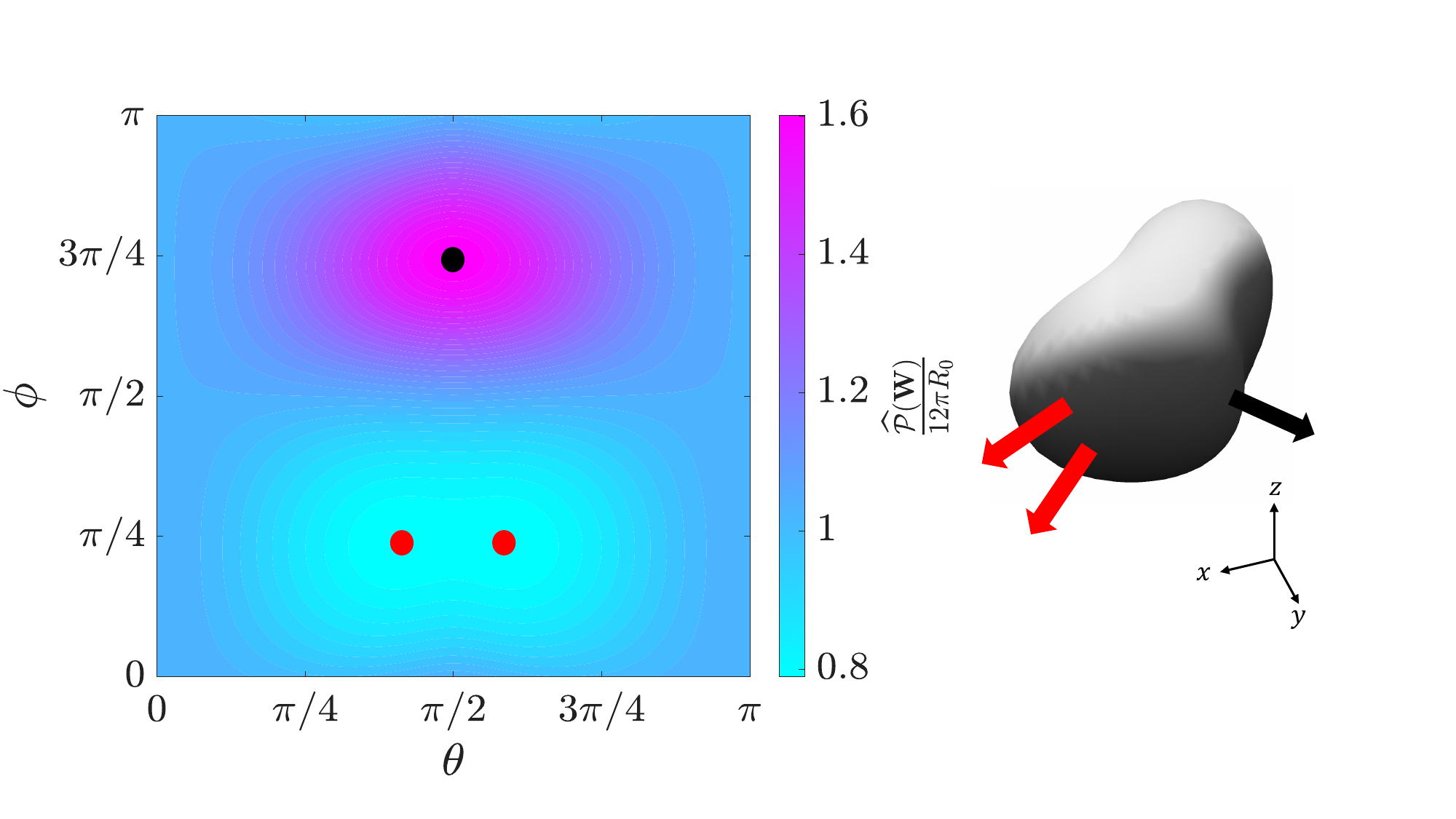}
    \caption{Contour plot of the partially minimized power loss $\widehat{\mathcal{P}}(\boldsymbol{W})$ as a function of the spherical coordinates $(\theta,\phi)$ when $\alpha = 0.3$ and $\beta = 0.2$. Two global minima are attained that are mirror reflections of each other in the $z = 0$ symmetry plane (red points). A global maximum is attained within the $z = 0$ plane (black point). The corresponding minimizing and maximizing net motion directions are drawn as red and black arrows, respectively. }
    \label{fig: asymmpowerlosslandscape}
\end{figure}

Figure \ref{fig: asymmpowerlosslandscape} shows the contour plot of $\widehat{\mathcal{P}}(\boldsymbol{W})$ when $\alpha = 0.3$ and $\beta = 0.2$. Making $\beta$ nonzero makes the lobes asymmetric, such that the shape loses its cyclic symmetry. While the global maximum located within the $z = 0$ plane remains, there are now two out-of-plane global minima that are mirror reflections of each other in the $z = 0$ symmetry plane. Since $\boldsymbol{W}_{\textrm{opt}}$ is no longer in a symmetry plane, globally optimal rotational motion is possible.

\begin{figure}
    \centering
    \includegraphics[trim=0 30 60 30, clip, width=0.9\textwidth]{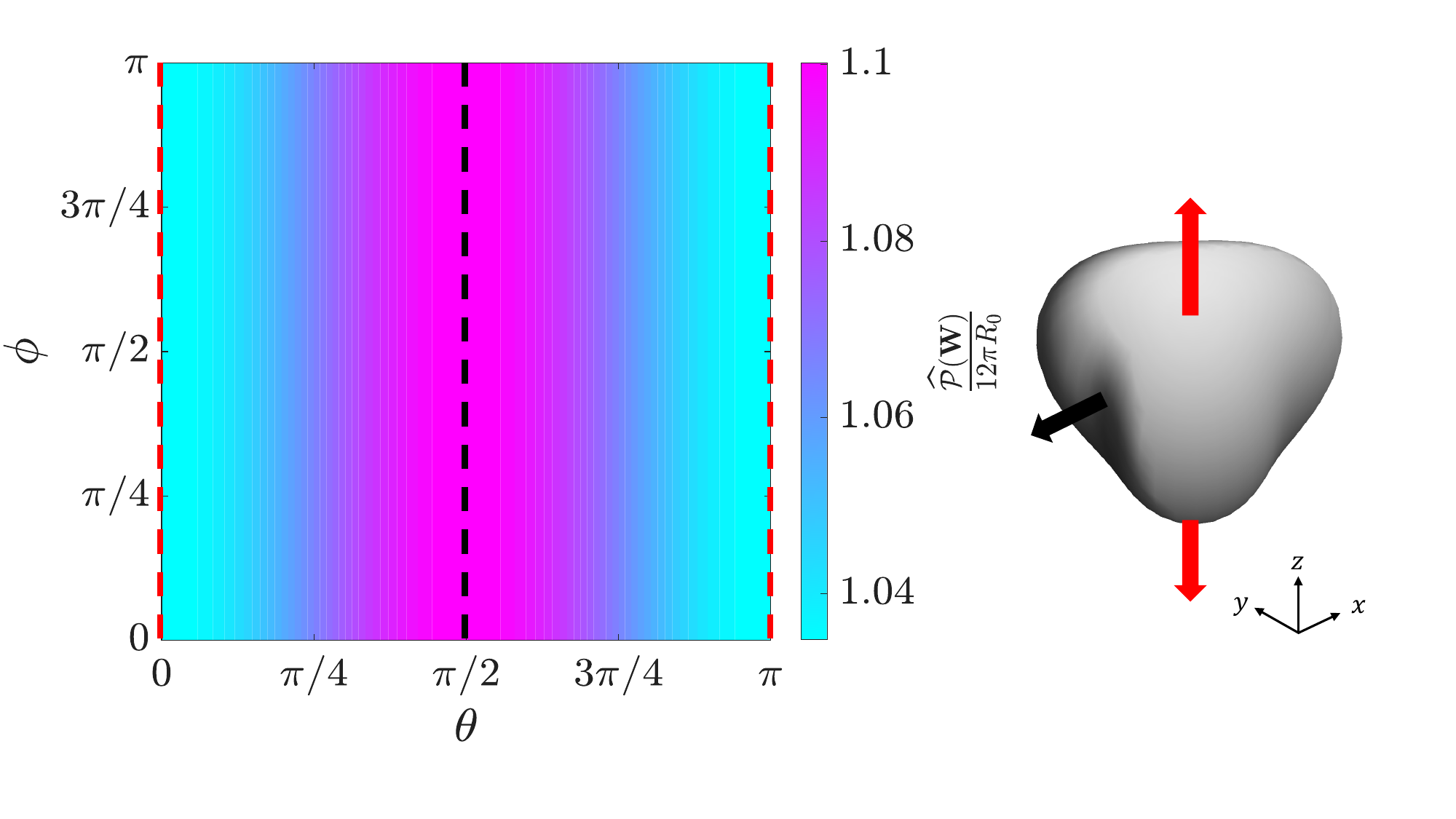}
    \caption{Contour plot of the partially minimized power loss $\widehat{\mathcal{P}}(\boldsymbol{W})$ as a function of the spherical coordinates $(\theta,\phi)$ when $\alpha = 0$ and $\beta = 0.2$. $\widehat{\mathcal{P}}(\boldsymbol{W})$ is independent of $\phi$ -- two global minima are attained at $\theta = 0$ and $\theta = \pi$ (red points) and any net motion direction in the $z = 0$ plane is a global maximizer (black dotted line). The corresponding minimizing directions $\boldsymbol{W}_{\textrm{opt}} = \pm \boldsymbol{e}_z$ are drawn as red arrows, and one possible maximizing direction is drawn as a black arrow. 
    }
    \label{fig: threelobepowerloss}
\end{figure}

Finally, we decrease $\alpha$ such that $\alpha = 0$ and $\beta = 0.2$, giving a shape with three symmetric lobes. In fact, this shape has a very high degree of symmetry -- both $z = 0$ and $y = 0$ are symmetry planes and the shape possesses order-3 cyclic symmetry as well. 
The contour lines in the contour plot in figure \ref{fig: threelobepowerloss} are now vertical, indicating that $\widehat{\mathcal{P}}(\boldsymbol{W})$ is independent of $\phi$. The two  global minima are pushed to the boundaries $\theta = 0$ and $\theta = \pi$, corresponding to $\boldsymbol{W}_\textrm{opt} = \pm \boldsymbol{e}_z$. Since $\boldsymbol{W}_\textrm{opt}$ is located within the $y = 0$ symmetry plane, the optimal motion is once again rotation-free. There are infinitely many maximizers, all located in the $z = 0$ plane (i.e. $\theta = \pi/2$). 

\begin{figure}
    \centering
    \includegraphics[trim=0 50 0 50, clip, width=\textwidth]{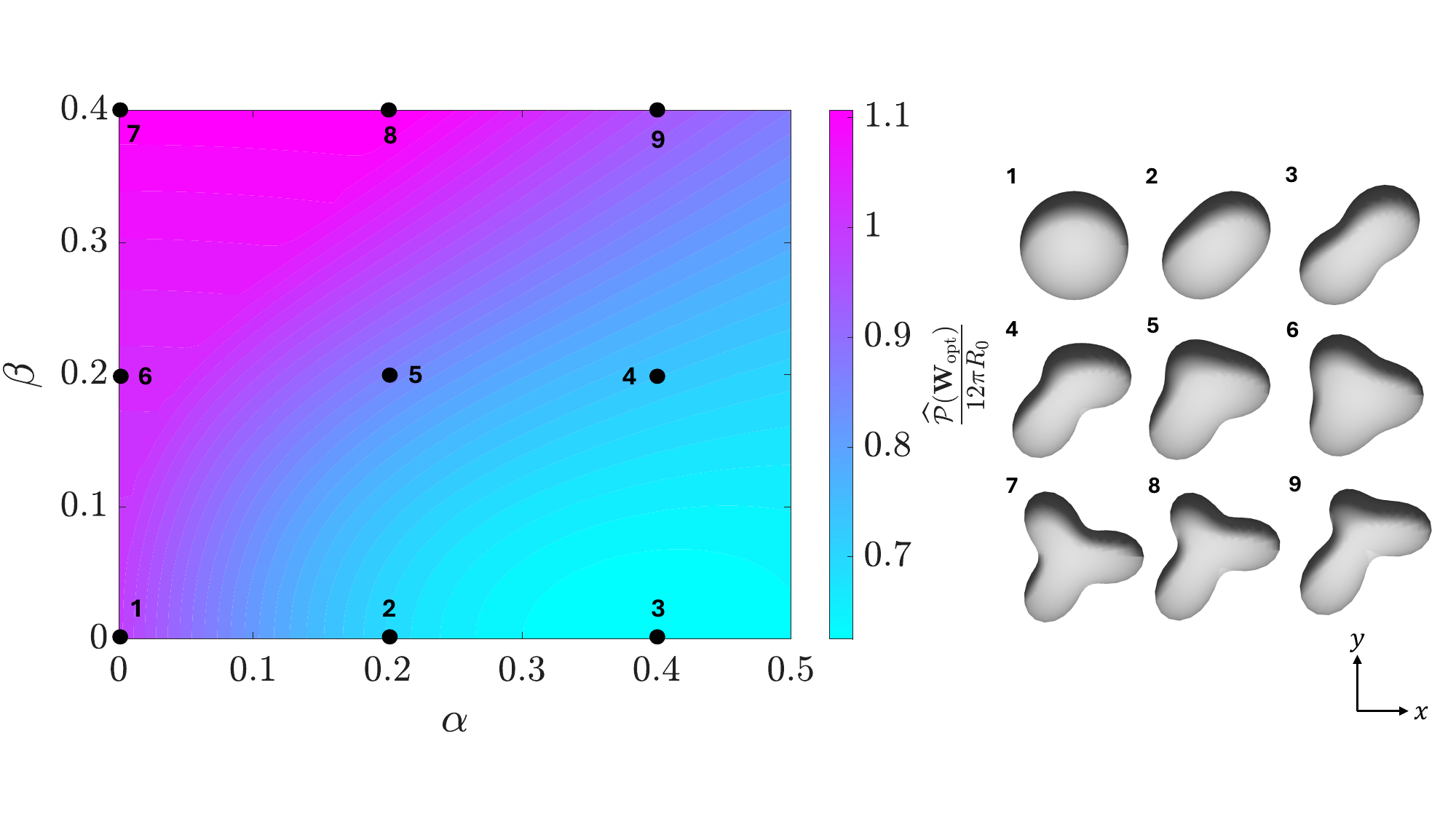}
    \caption{Globally optimal power loss $\widehat{\mathcal{P}}(\boldsymbol{W}_\textrm{opt})$ as a function of $\alpha \in [0, 0.5]$ and $\beta \in [0,0.4]$. The global optimization of Shape 3 given by $\alpha \approx 0.4$ and $\beta = 0$ produces the lowest power loss.}
    \label{fig: boomerangglobaloptpowerloss}
\end{figure}

Now that we have a better picture of the global optimization landscape of $\widehat{\mathcal{P}}(\boldsymbol{W})$ for various shapes in the family, we systematically vary $\alpha$ and $\beta$ and perform the global optimization for each shape to determine which shape can move with the least power loss. Figure \ref{fig: boomerangglobaloptpowerloss} shows a contour plot of the globally optimal power loss $\widehat{\mathcal{P}}(\boldsymbol{W}_{\textrm{opt}})$ as a function of $\alpha \in [0, 0.5]$ and $\beta \in [0, 0.4]$. The plot reveals that the symmetric dumbbell shape with $\alpha \approx 0.4$ and $\beta = 0$ achieves the least power loss. 

\begin{figure}
    \centering
    \includegraphics[trim=0 50 0 50, clip, width=\textwidth]{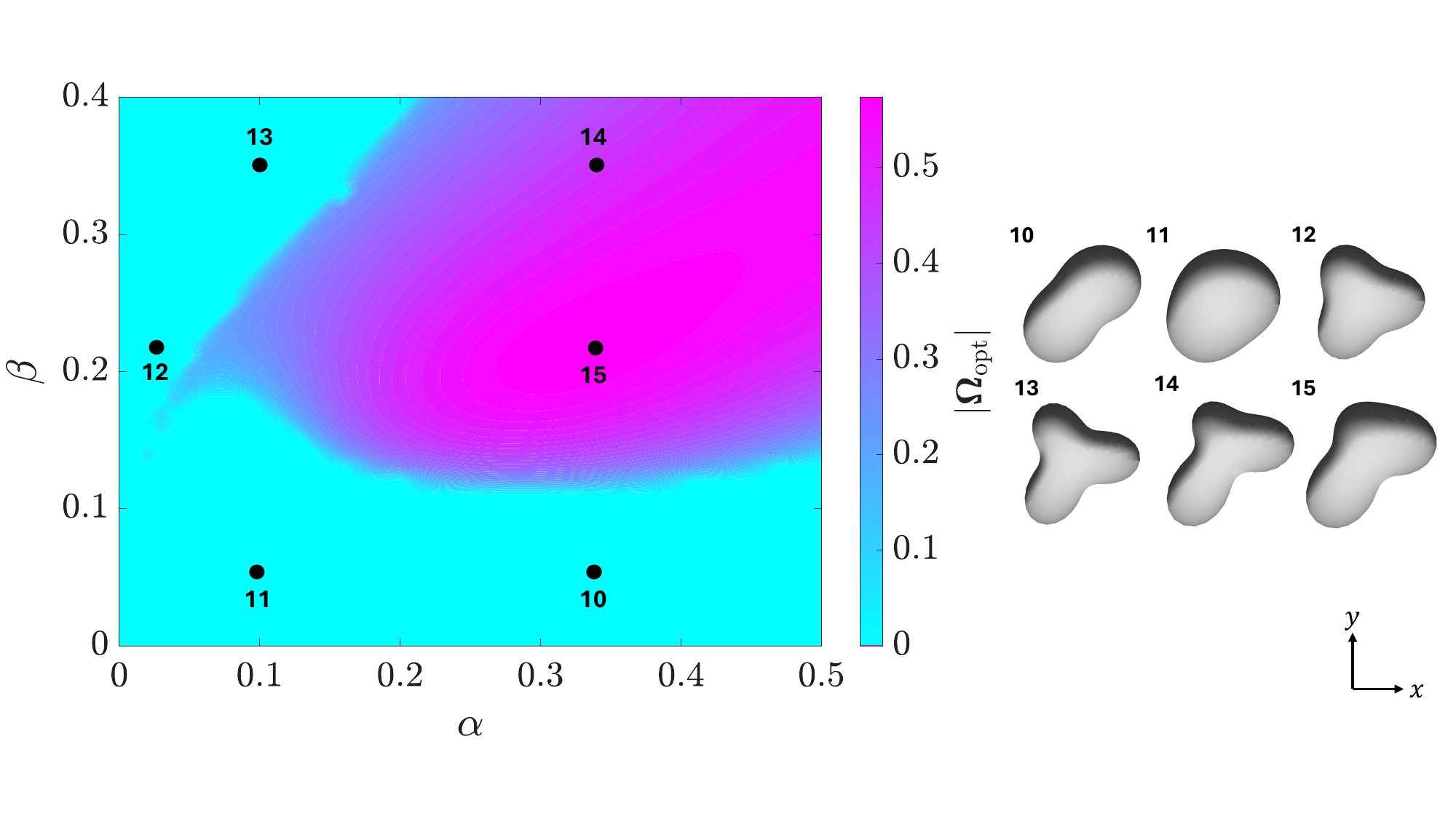}
    \caption{Globally minimizing angular speed $|\boldsymbol{\Omega}_{\textrm{opt}}|$ as a function of $\alpha \in [0, 0.5]$ and $\beta \in [0,0.4]$. The global optimization of Shape 15 given by $\alpha \approx 0.33$ and $\beta \approx 0.22$ produces the largest magnitude of rotation.}
    \label{fig: boomerangglobaloptomega}
\end{figure}

By plotting $|\boldsymbol{\Omega}_\textrm{opt}|$ as a function of $\alpha$ and $\beta$ (figure \ref{fig: boomerangglobaloptomega}), we can determine the shapes for which the globally optimal motion is rotational. When either $\alpha$ or $\beta$ is small, the shape is similar in form to the symmetric two and three-lobed shapes studied in figures \ref{fig: twolobecontour} and \ref{fig: threelobepowerloss} -- thus $\boldsymbol{\Omega}_{\textrm{opt}} \approx \boldsymbol{0}$. However, once $\alpha$ and $\beta$ are sufficiently large, the lobes become sufficiently asymmetric and $\boldsymbol{\Omega}_{\textrm{opt}} \neq \boldsymbol{0}$ for these shapes.

\section{Conclusions}
\label{sc:conclusion}

In this work, we have developed a framework to define the slip velocity on squirmers of arbitrary shape using the Helmholtz decomposition and spherical harmonic tangential basis functions. We proved that the trajectory followed by a steady squirmer of arbitrary shape is a circular helix;  in degenerate cases, the helix can take the form of a point, straight line, or circle. The rigid body velocities that define the helical trajectory can be obtained from the prescribed slip velocity by solving a simple six-dimensional linear system.

To better understand how geometry affects the rigid body velocities, we obtain analytical expressions for the rigid body velocities in terms of first-order squirming modes for a prolate spheroid. The formulas reveal relationships between the aspect ratio of the spheroid and the rigid body velocity components. While $U_z,\Omega_x, \Omega_y,$ and $\Omega_z$ decrease in magnitude as the spheroid becomes more slender, $U_x$ and $U_y$ increased in magnitude. 

We then transitioned from the forward problem to the inverse problem of finding the slip velocity that minimizes power loss given an arbitrary shape. The optimization problem can be decomposed into two nested minimization problems -- the first is a partial minimization problem in which the direction of net motion is prescribed. When the direction of net motion is along the axis of rotation for an axisymmetric shape, the optimal motion is rotation-free, in agreement with the result from \citet{stone1996propulsion} that rotation is inefficient for axisymmetric swimmers. This observation also holds for non-axisymmetric shapes when the prescribed net motion direction is within a plane of symmetry of a shape. However, when the net motion direction does not lie within a symmetry plane, rotational motion can reduce the total power loss. 

While the partial minimization prescribes a direction of net motion, we can subsequently perform a global minimization in which the best direction of net motion is determined. For some shapes, the optimal direction lies in a symmetry plane, leading to no rotation. However, we were able to find a simple shape for which the global optimization gave a rotational solution and the optimal direction was not located within a symmetry plane. Hence, for some shapes it can be advantageous to rotate even when the swimmer is allowed to move in any net direction. We generalized this result to a larger family of shapes and determined that asymmetry in the lobes of the shape leads to globally optimal rotational motion. Further work is needed to gain physical intuition for the global optimization problem -- unlike the partial optimization problem, an analytical solution is not known. 

Additionally, it would be useful to classify squirmers of arbitrary shape as pushers and pullers based on the squirming modes in our slip decomposition. 
It is important to note that while our slip velocity basis functions are highly general, more convenient basis choices exist for certain shapes. For example, \citet{poehnl2020axisymmetric} use prolate coordinates rather than spherical coordinates to study spheroidal squirmers. 

While our work thus far has focused on squirmers in free space, we have not studied arbitrary-shape swimmers in confining geometries such as pipes and shells, or near walls. The helical trajectories observed in Section \ref{sc:helix} will likely take a different form in these new geometries. Researchers have made progress studying ellipsoidal and axisymmetric squirmers near boundaries, such as \cite{ishimoto2013squirmer}, but studies of non-axisymmetric squirmers near these boundaries are still open. Our theoretical and computational framework for these arbitrary squirmers can help tackle this question. 

In addition to understanding the effect of constrained geometries, collective behavior of squirmer suspensions is of great interest. \cite{maity2022near} studied the interaction of two spherical chiral squirmers and \cite{samatas2023hydrodynamic} examined bulk suspensions of chiral squirmers using the lattice Boltzmann method. It would be helpful to explore how inhomogeneities in particle shape can affect the collective behavior of these suspensions by studying the interactions between squirmers of different shapes. Recent advances in boundary integral equations for swimmer suspensions can help reduce the computational complexity that comes with these large scale problems \citep{yan2020scalable}.


\begin{bmhead}[Funding]
KD gratefully acknowledges support from NSF GRFP under grant DGE-2241144. SV acknowledges support from NSF under grant DMS-2513346.
\end{bmhead}

\begin{bmhead}[Declaration of interests]
The authors report no conflict of interest.
\end{bmhead}



\begin{appendix}
\section{Shape equations}
\label{ap: shape gallery}
The analytical expressions for the radial profiles $\rho(\theta,\phi)$ and surface parameterizations $\boldsymbol{x}(\theta,\phi)$ for the shapes used in this work are detailed in Table \ref{tab:shape_gallery}.
\begin{table}
  \centering
  \begin{tabular}{ll}
    \textbf{Shape} & $\rho(\theta,\phi)$ \\
    \hline
    triangular antiprism & $ 1 + 0.2 \exp(-3 \textrm{Re}(Y_3^2(\theta,\phi)))$ \\[6pt]
    tilted dumbbell      & $ 1 + \textrm{Re}(Y_2^1(\theta,\phi)) + 0.1 \textrm{Re}(Y_3^2(\theta,\phi))$\\[6pt]
    \hline
    & $\boldsymbol{x}(\theta,\phi)$ \\ 
    \hline
    ellipsoid   & $(a \sin \theta \cos \phi, b \sin \theta \sin \phi, c \cos \theta)$ \\[6pt]
    butternut   & $((1-c\cos(\frac{\pi}{2}(\cos \theta - 0.2)))\sin \theta \cos \phi, (1-c\cos(\frac{\pi}{2}(\cos \theta - 0.2)))\sin \theta \sin \phi,\cos \theta)$ \\[6pt]
    stomatocyte & $(1.5+ \cos \theta)(\sin(\lambda \pi \sin \theta) \cos \phi,\sin(\lambda \pi \sin \theta) \sin \phi,\cos(\lambda \pi \sin \theta)) - 0.5 \boldsymbol{e}_z$ \\[6pt]
   \end{tabular}
  \caption{Geometric definitions for the particle shape gallery.}
  \label{tab:shape_gallery}
\end{table}

\end{appendix}







\bibliographystyle{jfm}
\bibliography{refs}

\end{document}